\documentclass{aa}  

\usepackage{graphicx}
\usepackage[varg]{txfonts}
\usepackage{natbib}
\bibpunct{(}{)}{;}{a}{}{,} 

\newcommand\mone{$^{-1}$}
\newcommand\mtwo{$^{-2}$}

\newcommand\tcd{\object{3C~293}}
\newcommand\cosc{\element[][]{CO(6-5)}}
\newcommand\coct{\element[][]{CO(4-3)}}
\newcommand\cotd{\element[][]{CO(3-2)}}
\newcommand\codu{\element[][12]{CO(2-1)}}
\newcommand\couc{\element[][12]{CO(1-0)}}

\newcommand\msun{M$_{\sun}$}
\newcommand\lsun{L$_{\sun}$}
\newcommand\hd{H$_2$}
\newcommand\mhd{$M$(H$_2$)}
\newcommand\hi{\ion{H}{i}}
\newcommand\vs{$v_{\rm{sys}}$}
\newcommand\vsco{$v_{\rm{sys}}^{\rm{CO}}$}
\newcommand\vc{$V_{\rm{circ}}$}
\newcommand\vnc{$V_{\rm{nc}}$}
\newcommand\msyr{\msun\ yr\mone}
\newcommand\ssfr{$\Sigma_{\rm{SFR}}$}
\newcommand\shd{$\Sigma_{\rm{MH2}}$}
\newcommand\kms{km~s\mone}
\newcommand\mum{$\mu$m}
\newcommand\grados{$^\circ$}
\newcommand\so{S$_{8\mu\mathrm{m,dust}}$}
\newcommand\kin{{\it kinemetry}}
\usepackage{color}

\begin{document}

\title{Fueling the central engine of radio galaxies}
\subtitle{III. Molecular gas and star formation efficiency of \tcd. } 
\titlerunning{The molecular gas and SFE of \tcd}
\author{
A. Labiano\inst{1}
\and
S. Garc\'ia-Burillo\inst{2}
\and
F. Combes\inst{3}
\and 
A. Usero\inst{2}
\and 
R. Soria-Ruiz\inst{2}
\and \\
J. Piqueras L\'opez\inst{1}
\and 
A. Fuente\inst{2}
\and
L. Hunt\inst{4}
\and
R. Neri\inst{5}
}
\offprints{\\ Alvaro Labiano: {\tt labianooa@cab.inta-csic.es } \smallskip \\
{\it Based on observations carried out with the IRAM Plateau de Bure Interferometer. IRAM is supported by INSU/CNRS (France), MPG (Germany) and IGN (Spain). }}

\institute{
Centro de Astrobiolog\'ia (CSIC-INTA), Carretera de Ajalvir km. 4, 28850 Torrej\'on de Ardoz, Madrid, Spain.
\and
Observatorio Astron\'omico Nacional, Alfonso XII, 3, 28014, Madrid, Spain.
\and
Observatoire de Paris, LERMA \& CNRS: UMR8112, 61 Av. de l'Observatoire, 75014 Paris, France.
\and
INAF/Osservatorio Astrofisico di Arcetri, Largo Enrico Fermi 5, 50125 Florence, Italy.
\and
IRAM, 300 rue de la Piscine, Domaine Universitaire, 38406 St. Martin d'H\'eres Cedex, France.
}

\date{  }

\abstract
{Powerful radio galaxies show evidence of ongoing active galactic nuclei (AGN) feedback, mainly in the form of fast, massive outflows. But it is not clear how these outflows affect the star formation of their hosts.
}
{
We investigate the different manifestations of AGN feedback in the evolved, powerful radio source \tcd\  and their impact on the molecular gas of its host galaxy, which harbors young star-forming regions and fast outflows of \hi\ and ionized gas.
} 
{
We study the distribution and kinematics of the molecular gas of \tcd\  using high spatial resolution observations of the \couc\ and \codu\ lines, and the 3~mm and 1~continuum taken with the IRAM Plateau de Bure interferometer. We mapped the molecular gas of \tcd\ and compared it with the dust and star-formation images of the host.  We searched for signatures of outflow motions in the CO kinematics, and re-examined the evidence of outflowing gas in the \hi\ spectra. We also derived the star formation rate (SFR) and  star formation efficiency (SFE) of the host with all available SFR tracers from the literature, and compared them with the SFE of young and evolved radio galaxies and normal star-forming galaxies.
}
{The \couc\ emission line shows that the molecular gas in \tcd\ is distributed along a massive (\mhd$\sim$2.2$\times$10$^{10}$~\msun)  $\sim$24$\arcsec$(21~kpc-) diameter warped disk, that rotates around the AGN. 
Our data show that the dust and the star formation are clearly associated with the CO disk.  The \codu\ emission is located in the inner 7~kpc (diameter) region around the AGN, coincident with the inner part of the \couc\ disk. Both the \couc\ and \codu\ spectra reveal the presence of an absorber against the central regions of \tcd\ that is associated with the disk.  We do not detect any fast ($\gtrsim$500~\kms) outflow motions in the {  cold molecular gas}.
 The host of \tcd\ shows an SFE consistent with the Kennicutt-Schmidt law of normal galaxies and young radio galaxies, and it is 10-50 times higher than the SFE estimated with the 7.7~\mum\ PAH emission of evolved radio galaxies. Our results suggest that the apparently low SFE of evolved radio galaxies may be caused by an underestimation of the SFR and/or an overestimation of the molecular gas densities in these sources.
}  
{
The molecular gas of \tcd, while not incompatible with a mild AGN-triggered flow, does not reach the high velocities ($\gtrsim$500 \kms) observed in the \hi\ spectrum. We find no signatures of AGN feedback in the molecular gas of \tcd.
}


\maketitle
%

\section{Introduction}


An active galactic nucleus (AGN) releases vast amounts of energy into the interstellar medium (ISM) of its host galaxy. This energy transfer, known as AGN feedback, may increase the temperature of the ISM gas, preventing its collapse, and/or expel it in the form of outflows {  \citep[see][ for a review]{Fabian12}}. Both effects will inhibit the star formation of the host and affect its evolution.
Observational evidence of the effects of AGN feedback on the hosts is frequently found in the literature \citep{Thomas05, Murray05, Schawinski06, Muller06, Feruglio10, Crenshaw10, Fischer11, Villar11, Dasyra11, Sturm11, Maiolino12, Aalto12, Walter13, Silk13}. It is thought that AGN feedback might be responsible for several properties observed in galaxies, such as the correlations between black hole and host galaxy bulge mass \citep{Magorrian98, Tremaine02, Marconi03, Haring04}, or the fast transition of early-type galaxies from the blue cloud to the red sequence \citep{Schawinski07, Kaviraj11}. 
Hence, it is now clear that models of galaxy evolution need to consider the effects of AGN feedback in the host galaxy \citep{King03, Granato04, Matteo05, Croton06, Ciotti07, Menci08, King08, Merloni08, Narayanan08, Silk10}.


The AGN feedback may take place through two mechanisms: through
AGN radiation (known as radiative or quasar mode of AGN feedback), or through interaction of the jet with the ISM (known as kinetic or radio mode).
The most intense effects of the kinetic mode are seen in powerful radio galaxies. The extremely large and fast jets of powerful radio galaxies are able to inject enormous amounts of energy into their host galaxy ISM and even into the intergalactic medium \citep[IGM; e.g.,][]{Birzan04,Fabian06,McNamara07,McNamara13,Russell13}. Most massive galaxies will harbor a radio galaxy phase during their lifetime \citep[e.g.,][]{Best06}. Studying AGN feedback during this phase is thus crucial for understanding galaxy evolution.

\begin{figure*}
\centering
\includegraphics[width=1.5\columnwidth,angle=0]{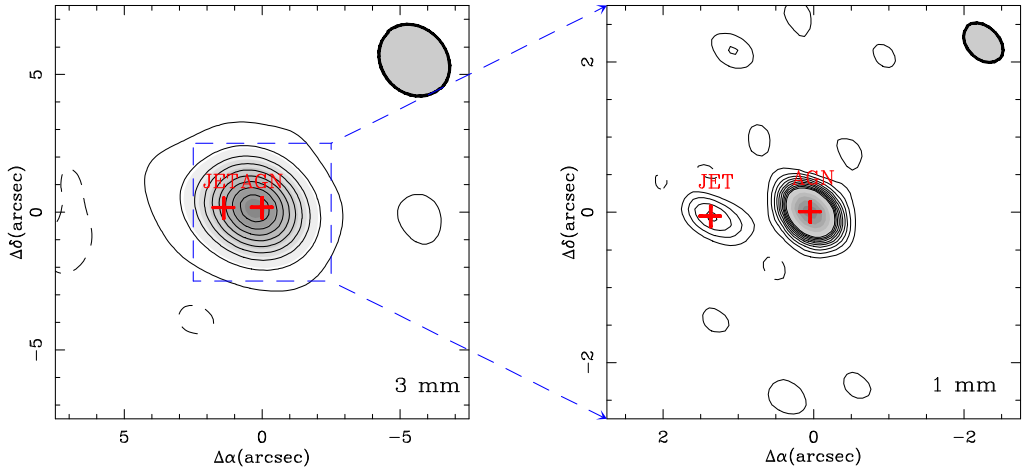} 
\caption{Continuum maps of \tcd\ at 3~mm (left) and 1~mm (right). The positions of the AGN and jet components, based on the UV-FIT results (Table \ref{uvfits}), are identified with crosses (+).
 The dashed box marks the size of the 1 mm map shown in the right panel. The contour levels are  --9, 9, 39 to 249 mJy~beam\mone, in steps of 30 mJy~beam\mone\  for the 3~mm map; and --6, 6, 12 to 60 mJy~beam\mone, in steps of 6 mJy~beam\mone\ for the 1~mm map. 
  ($\Delta\,\alpha$, $\Delta\,\delta$)=(0,0) corresponds to the position of the AGN at 1 mm. The gray ellipses in the top-right corner show the PdBI beam for each map. \label{contfits}}
\end{figure*}

A large number of radio galaxies show manifestations of ongoing AGN feedback at different redshift
ranges, whether as gas outflows or as inhibition of star formation. Outflows in radio galaxies have been detected through \hi\ and ionized gas observations \citep[e.g.,][]{Morganti03c, Rupke05, Holt06, Nesvadba06, Nesvadba08, Lehnert11}. Searches for molecular gas, which may dominate the mass/energy budget of the wind, show evidence of outflows only in one radio source  \citep[4C~12.50;][]{Dasyra11}.
Based on the molecular gas contents, and SFR estimations using the 7.7 $\mu$m polycyclic aromatic hydrocarbons (PAH) emission, \citet{Nesvadba10} found that radio galaxies can be $\sim$10--50 times less efficient in forming stars than galaxies that follow the canonical Kennicutt-Schmidt \citep[KS,][]{Schmidt59, Kennicutt98} relationship. The warm-\hd\ emission of these radio galaxies suggests shocks in the ISM that are induced by the expansion of the powerful radio jet. These shocks are thought to increase the turbulence in the molecular gas, which inhibits the star formation in the host galaxy. However, it is not clear how jet-induced shocks affect the cold phase of the molecular gas. 

Our recent work on \object{3C~236} \citep{Labiano13} found that not all radio galaxies show a low star formation efficiency (SFE). \object{3C~236} is a re-activated radio source that shows an SFE consistent with the KS-law. {  The young age of its second period of  radio-source activity \citep[$10^5$ yr, based on the dynamics of the source,][]{Tremblay10,O'Dea01}, and the older age of the \citet{Nesvadba10} radio source sample ($10^7-10^8$ yr) } suggested that the impact of the AGN feedback on the star
formation rate probably evolves with time. A literature search showed that different authors had measured the SFR and \mhd\ of the young sources: \object{4C~12.50}, \object{4C~31.04} \citep{Burillo07,Willett10, Dasyra12}. These two sources show an SFE consistent with the KS-law, which supports the evolutionary scenario of the feedback impact. 
Observations of cold molecular gas in the ISM of different populations of young and evolved radio galaxies and accurate measurements of their SFR are the key to understanding how AGN feedback affects the properties of their hosts.

\subsection{\tcd}

The radio source \tcd\ is a large ($\sim$200 kpc), powerful ($P_{\rm 5GHz}$=9$\times$$10^{24}$ W~Hz\mone), 
 FR II\footnote{\tcd\ also shows properties of an FR I radio galaxy. In the literature, it is sometimes classified as an FR I or intermediate FR I/II radio galaxy \citep[e.g.,][]{Leipski09, Massaro10, Tadhunter11}.}  \citep{Fanaroff74} radio galaxy with PA=135\grados. The inner 2$\arcsec$ of the radio source show a compact core and two jets aligned east-west (PA=93\grados), with the eastern jet approaching. 
Estimations based on spectral aging show that the large radio source was triggered $\lesssim$20 Myr ago and stopped $\sim$0.35 Myr ago, while the compact source was triggered $\lesssim$0.18 Myr ago \citep{Akujor96, Joshi11}. 
The time-lapse between the two activity epochs probably lasted only  $\lesssim$0.1 Myr \citep{Joshi11}.
This short pause in activity implies that the radio jets of \tcd\ may have been almost continuously interacting with the ISM of the host for the last $\sim$20 Myr.

The host galaxy of \tcd\ \citep[\object{UGC~8782} or \object{VV5-33-12},  $z$=0.0450,][]{Vaucouleurs91} has been classified both as a spiral galaxy \citep[e.g.,][]{Sandage66, Colla75, Burbidge79} and as an elliptical galaxy \citep[e.g.,][]{Veron01, Tremblay07}, probably because of its extremely complex morphology, with a dust disk, compact knots, and large dust lanes\footnote{The total mass of dust in \object{UGC~8782} is $\sim$$10^7$ \msun\ \citep{Koff00, Papadopoulos10}.}. The morphology and the vast amounts of gas and dust found in the host of \tcd\ are consistent with a gas-rich merger of an elliptical galaxy with a spiral galaxy \citep{Martel99, Koff00, Capetti00, Floyd06}. \object{UGC~8782} shows an optical emission tail that extends beyond a possible companion $\sim$37$\arcsec$ to the southwest \citep{Heckman86, Smith89}. It is not clear whether this companion galaxy is interacting with \object{UGC~8782} or was formed in the dust tail \citep{Floyd06}. 

Hubble Space Telescope (HST) near-UV (NUV) photometry of \object{UGC~8782} revealed several regions of bright emission along the edges of the dust lanes, consistent with recent star formation  \citep{Allen08, Baldi08}. Optical spectroscopy shows that the total stellar mass of \tcd\ is 2.8$\times$10$^{11}$ \msun, consisting of a young (0.1-2.5 Gyr) stellar population, which represents 57\% of the total stellar mass, and an old stellar population \citep[$\gtrsim$10 Gyr,][]{Tadhunter05}. The age of the young stellar population is consistent with a star formation episode after the merger \citep{Tadhunter05, Tadhunter11}. The {\it Starlight} \citep{Cid05} fit of the SDSS spectrum Ca H+K lines showed that the stellar population of \tcd\ can be described by three components with ages 10 Gyr, 0.1--1 Gyr, and 10 Myr \citep{Nesvadba10}.  

The \hi\ spectrum of \tcd\ shows a broad ($\sim$1200 \kms)  absorption component associated with the western jet of the compact radio source (0.5 kpc west of the nucleus). This absorption is created in an outflow of neutral gas  produced by the interaction of the radio jet with the ISM \citep{Morganti03, Morganti05, Mahony13}. Optical, long-slit spectra show an outflowing component in the ionized gas emission lines, with a width and center consistent with those of the \hi\ outflow. This ionized gas outflow, however, is associated with the east component of the compact radio source (1 kpc east of the nucleus). Weaker hints of outflowing ionized gas might also be present in the center and western components of the compact radio source \citep{Emonts05}. 


Based on CO observations made with the NRAO-12m telescope and the OVRO Millimeter Array, \citet{Evans99b} reported the detection of the \couc\  line, both in emission and absorption, in \tcd\  \citep[see also][]{Evans05}. According to their data, which
have a spatial resolution of  $\sim$3.5$\arcsec$, the \couc\ emission extends across a 7$\arcsec$ diameter disk, with a total mass of cold \hd\ of  \mhd$\sim$1.5$\times$10$^{10}$ \msun, and is perpendicular to the large-scale radio source. 
 Later observations, using the JCMT, detected high-excitation CO emission lines in \tcd\ \citep[\coct, \cosc,][]{Papadopoulos08, Papadopoulos10}. 
The PdBI was also used to probe the molecular gas using the \codu\  line, and search for evidence of outflow \citep{Burillo09}. None of the observations above had the required velocity coverage and sensitivity to detect the outflow seen in \hi\ and ionized gas, however.  

Even though \tcd\ has large contents of molecular gas and young stars, \citet{Nesvadba10} measured an SFE 20 times lower than expected for normal star-forming galaxies, based on its 7.7 \mum\ PAH emission and the \hd\ measurements of \citet{Evans99b}. They attributed this low SFE to the impact of AGN feedback on the ISM through the outflows observed in \hi\ and ionized gas. On the other hand, the FIR luminosity of \tcd\  \citep[$3.3\times10^{10}$~\lsun,][]{Floyd06} is consistent with FIR luminosities of normal, KS-law, star-forming galaxies with similar molecular gas contents, suggesting that the SFR of \tcd\ calculated with the 7.7 \mum\ PAH emission may be underestimated. 

The large molecular gas content of the host, the ages of the old radio source and young stellar population, the re-ignition of the nuclear activity, and  gas outflows caused by jet-ISM interactions make \tcd\ an ideal candidate for studying the impact of AGN feedback in radio galaxies.
We carried out 1~mm, 3~mm continuum and \couc, \codu\ line observations of \tcd\ with the Plateau de Bure Interferometer (PdBI)  to investigate the impact of AGN feedback of activity in its host. Our sensitivity and resolution at 3 mm are, respectively, five  and two times  better than achieved before \citep[e.g.,][]{Evans99b},
 and we covered a velocity range broad enough ($>$4000 \kms) to detect the molecular gas counterpart of the \hi\ and ionized gas outflows. 
In this paper, we present the results of these observations where we study the continuum emission of \tcd, and analyze the morphology and kinematics of the molecular gas. We determine the systemic velocity of \tcd, search for outflow signatures in the {  cold molecular gas}, and compare our results with the \hi\ observations of the outflow. We then reassess the SFR of \tcd\ based on several SFR tracers, taking care to examine possible effects of the AGN. Finally, we compare the SFE of \tcd\ with a sample of powerful radio galaxies and normal star-forming galaxies and examine the possible impact of AGN feedback on star formation.

\section{Observations}

\subsection{Interferometer observations}

We observed \tcd\ in the \couc\ and \codu\ lines, at 3 and 1 mm using six antennas in the D and B configurations, respectively, of the IRAM PdBI \citep[][]{Guilloteau92} in January 2011. The receivers were tuned based on the redshift $z$=0.0446 ($\varv_0$=13\,371~km~s\mone), so that the \couc\ and \codu\ lines were centered on 110.350 and 220.695 GHz, respectively. 
With this setting we obtained a velocity coverage of $\sim$10\,000~\kms\ at 110.350 GHz and 5\,000~\kms\ at 220.695 GHz, with the WideX wide-band correlator of the PdBI (3.6~GHz-wide) and the two polarizations of the receiver. Observations were conducted in a single pointing of FWHM $\sim$45$\arcsec$ at 110.350 GHz and $\sim$23$\arcsec$ at 220.695 GHz. The adopted phase-tracking center of the observations was set at ($\alpha_{2000}$, $\delta_{2000}$)=(13$^{\rm h}52^{\rm m}17.8^{\rm s}$, 31$^\circ$26$\arcmin$46.2$\arcsec$), the position of the nucleus given by the NASA/IPAC Extragalactic Database (NED). Nevertheless, {   the position of the AGN} determined in Sect. \ref{contmap}, which  coincides with the position of the radio continuum VLBI core \citep{Beswick04}, is $\simeq$0.3$\arcsec$ offset to the north with respect to the array center: ($\Delta\,\alpha$, $\Delta\,\delta$)$\sim$(0.0$\arcsec$, 0.3$\arcsec$). Hereafter, all positions are given with respect to the {  coordinates of the AGN}. Visibilities were obtained through on-source integration times of  22.5 min framed by short (2.25 min) phase and amplitude calibrations on the nearby quasar \object{1308+326}. The visibilities were calibrated using the antenna-based scheme. The absolute flux density scale was calibrated on \object{MWC349} and found to be accurate to $\la$10\% at 220.695 GHz. 

The image reconstruction was made with the standard IRAM/GILDAS software \citep{Guilloteau00}\footnote{http://www.iram.fr/IRAMFR/GILDAS}.  We obtained the maps of the continuum emission averaging channels free of line emission from $\varv$-$\varv_o$=--1810 to --3114\,km\,s$^{-1}$ for 110.350\,GHz, and $\varv$-$\varv_o$=--1758 to --3035\,km\,s$^{-1}$ for 220.695\,GHz. The corresponding 1$\sigma$ sensitivities of the continuum emission are $\sim$0.4\,mJy\,beam$^{-1}$ at 110.350\,GHz, and $\sim$0.7\,mJy\,beam$^{-1}$ at 220.695\,GHz. 
 We used natural weighting and no taper to generate the CO line maps, with a field-of-view (FoV) of 51$\arcsec$ and 0.2$\arcsec$/pixel sampling  and a synthesized beam of 2.4$\arcsec \times$2.8$\arcsec$\,@$PA$=42$\degr$ at 3~mm,  and a FoV  of 26$\arcsec$ with 0.1$\arcsec$/pixel sampling, and a synthesized beam of 0.44$\arcsec \times$0.61$\arcsec$\,@$PA$=44$\degr$ at 1~mm. The continuum levels were subtracted in the visibilities plane to generate the line maps. The 1$\sigma$ point source sensitivities were derived from emission-free channels resulting in 0.9\,mJy\,beam$^{-1}$ in 5.9\,MHz ($\sim$16\,km\,s$^{-1}$)-wide channels for 110.350\,GHz, and 1.7\,mJy\,beam$^{-1}$ in 5.9\,MHz ($\sim$8\,km\,s$^{-1}$)-wide channels for 220.695\,GHz.

\subsection{Supplementary data}

We used the following data from the HST archive:  STIS/NUV-MAMA/F25SRF2 \citep[hereafter NUV image,][]{Allen02, Floyd06}, WFPC2/F702W  \citep[$R$-band image,][]{Martel99, Koff00, Floyd06}, NICMOS2/F160W ($H$-band), F170M, F181M  and $R-H$ map from \citet{Floyd06}; \hi\ data by \citet{Morganti03}; mid-IR data from \citet{Dasyra11} and \citet{Guillard12}, and the Spitzer archive; Sloan Digital Sky Survey (SDSS) spectra from \citet{York00}, \citet{Abazajian09}, and \citet{Buttiglione09}; William Herschel Telescope (WHT) data from \citet{Emonts05}.

We use H$_0$=71, $\Omega_M$=0.27, and $\Omega_\Lambda$=0.73 \citep{Spergel03, Spergel07} throughout the paper. Luminosity and angular distances are D$_{L}$=196.9\,Mpc and D$_{A}$=180.3\,Mpc; the latter gives 1$\arcsec$= 0.874\,kpc \citep{Wright06}.

\begin{figure*}[]
\centering
\includegraphics{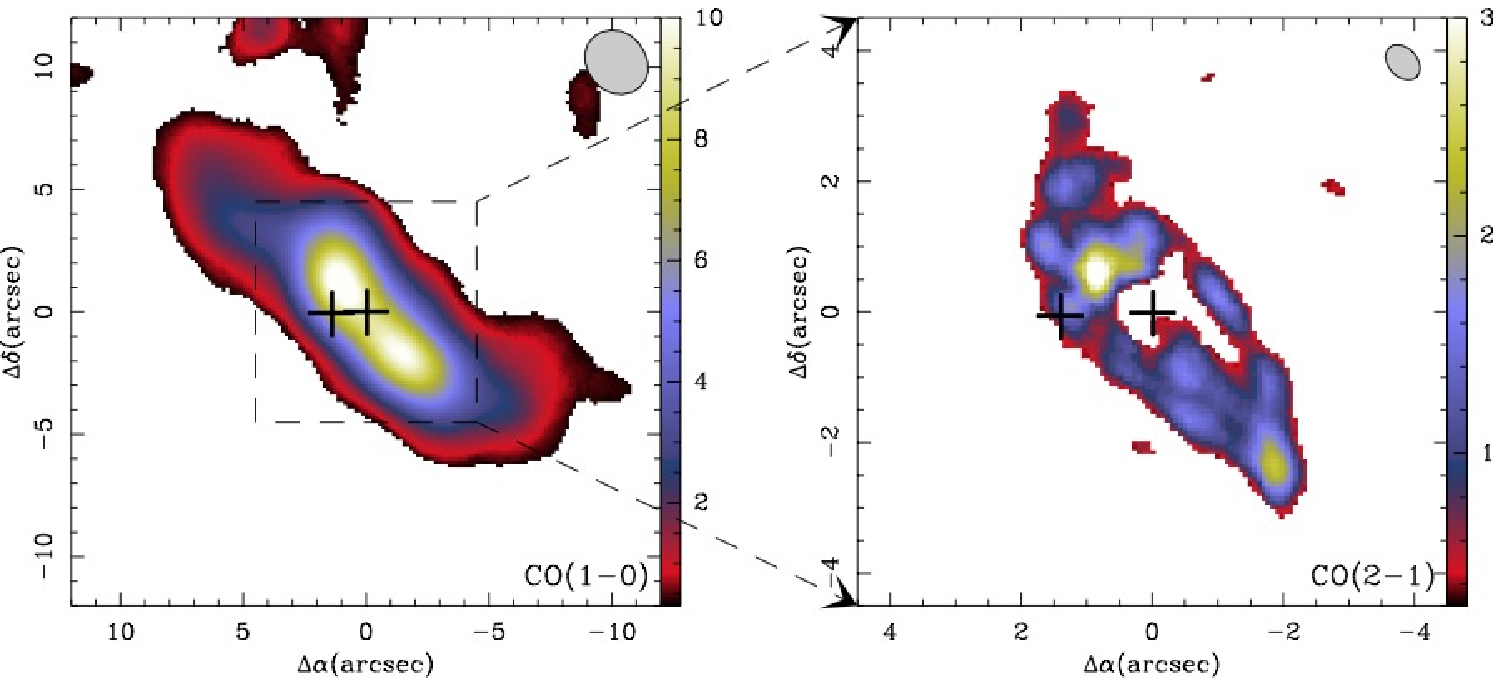}
\includegraphics{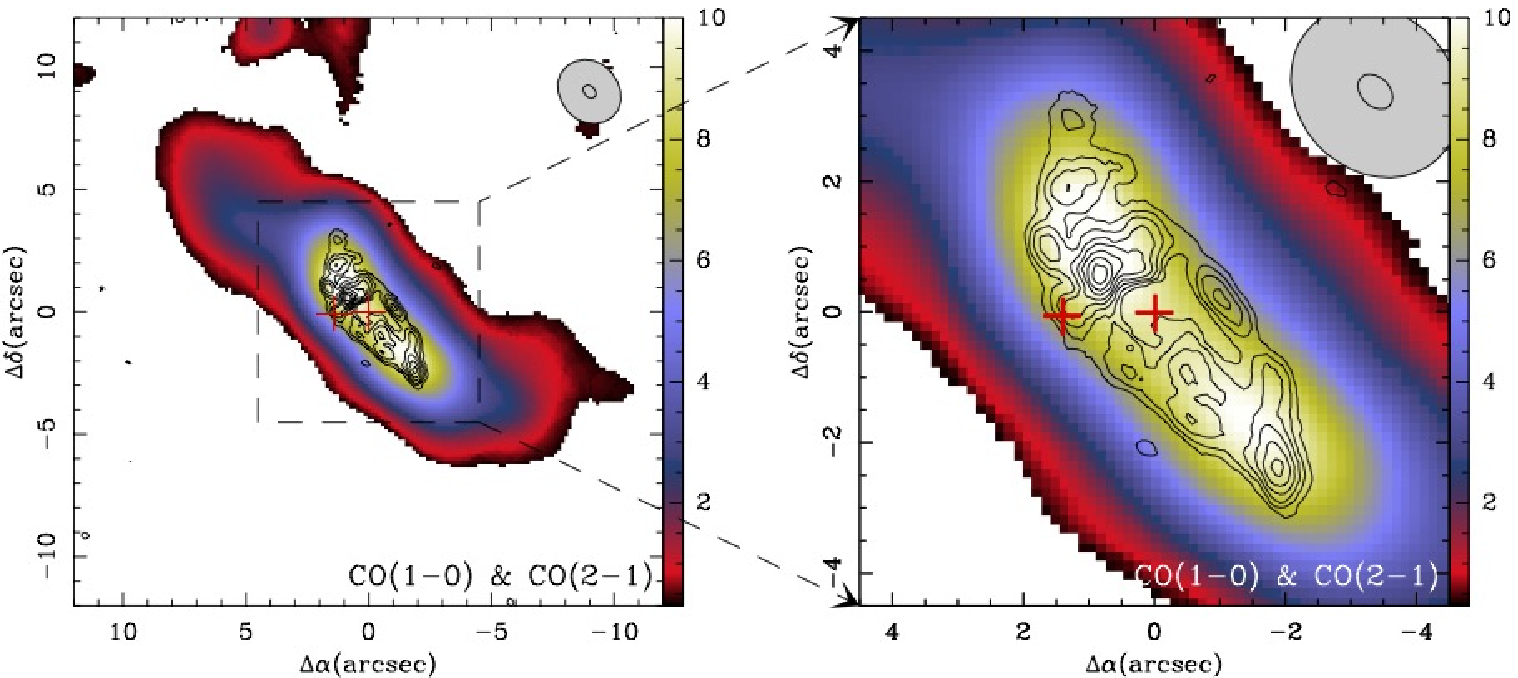}
\caption{\couc\ and \codu\ intensity maps of \tcd, integrating all emission above 3$\sigma$ levels, from  $v$-$v_o$=--380 to +500~\kms\ and $v$-$v_o$=--280 to +450~\kms\ respectively. Crosses (+) mark the position of the AGN and jet peak emission according to the UV-FIT models (Table \ref{uvfits}). The dashed box in the \couc\ map marks the size of the \codu\ map shown in the right panel. Color scales are in Jy beam\mone\ km s\mone. The maps were clipped at 3$\sigma$, with $\sigma$=0.11 and 0.13 ~Jy~beam\mone\ \kms\ for the \couc\ and \codu\ maps, respectively. 
The bottom panels show an overlay of  the \couc\ (color scale) and \codu\ (contours) emission above 3$\sigma$ levels. The contour levels are 0.39 to 3.9~Jy~beam\mone\ \kms, in steps of 0.39~Jy~beam\mone\ \kms.
\label{cointeg}}
\end{figure*}


\begin{table}
\caption{AGN and jet position.}
\label{uvfits}
\begin{minipage}{1.\columnwidth}
\resizebox{1.\textwidth}{!}{
\begin{tabular}{ccccc} 
\hline
\hline
 Component & Wavelength & RA (J2000) &  Dec  (J2000) &    Flux (mJy)  \\
\hline
AGN*   & 1~mm & 13:52:17.80  &  +31:26:46.460  &  183.4$\pm$0.9 \\
Jet   & 1~mm & 13:52:17.91  &  +31:26:46.410  &   28.9$\pm$0.9\\
AGN   & 3~mm & 13:52:17.80  &  +31:26:46.585  &  231$\pm$4  \\
Jet   & 3~mm & 13:52:17.91  &  +31:26:46.565  &   89$\pm$4  \\
\hline
VLBI core & 6~cm & 13:52:17.80 & +31:26:46.48 & --  \\
VLBI E1   & 6~cm & 13:52:17.91 & +31:26:46.48 & -- \\
\hline
\end{tabular}
}
\end{minipage}
\\ \\
{  AGN and jet positions in the 1 mm and 3 mm maps of \tcd, according to the UV-FIT models (point sources for the AGN and jet, with FWHM$_\mathrm{1~mm}\simeq0.52\arcsec$, FWHM$_\mathrm{3~mm}\simeq2.6\arcsec$). The VLBI positions of the core and jet component E1  \citep{Beswick04, Floyd06} are included for comparison.  The coordinates of the AGN at 1 mm (AGN*) are adopted as the AGN position throughout.}
\end{table}

\section{Continuum maps}
\label{contmap}

Figure~\ref{contfits} shows the 1 mm and 3 mm continuum maps of \tcd. 
The continuum emission at both wavelengths consists of a bright central component plus a fainter one $\sim$1$\farcs$3 to the east. Using the UV-FIT task of GILDAS, we fitted two point sources to the 1 mm and 3 mm continuum visibilities of \tcd\ {  (FWHM$_\mathrm{1~mm}\simeq0.52\arcsec$, FWHM$_\mathrm{3~mm}=2.6\arcsec$)}. The results of the best fits are given in Table~\ref{uvfits}. The central component, {  located at ($\Delta\alpha$,~$\Delta\delta$)$\sim$(0$\farcs$0,~0$\farcs$0), } is responsible for 86\% and 72\% of the emission at 1 mm and 3 mm, respectively. The  global-VLBI and MERLIN 1.35 and 1.5~GHz maps show that this central component corresponds to the position of the AGN \citep[component `Core'  in][]{Beswick04, Floyd06}. Therefore, we adopted this component as the {  AGN position} of \tcd\ (component labeled `AGN' in Fig. \ref{contfits}). 


The eastern component was fitted at both
wavelengths by a point source with coordinates  ($\Delta\alpha$,~$\Delta\delta$)$\sim$(+1$\farcs$3$\pm$0.1,~+0$\farcs$0$\pm$0$\farcs$1), consistent with the jet component E1 of the global-VLBI and MERLIN 1.35 and 1.5~GHz maps  \citep{Beswick04, Floyd06}. The coordinates of the fitted AGN and jet components yield a PA of 89\grados\ for the mm-jet(s), consistent with the orientation of the small radio jet determined in the VLBI cm-maps of the source. Therefore, we have detected the mm-counterparts of the approaching jet of the compact radio source of \tcd\ (component labeled `jet' in Fig. \ref{contfits}). 



\begin{figure*}
\centering
\includegraphics{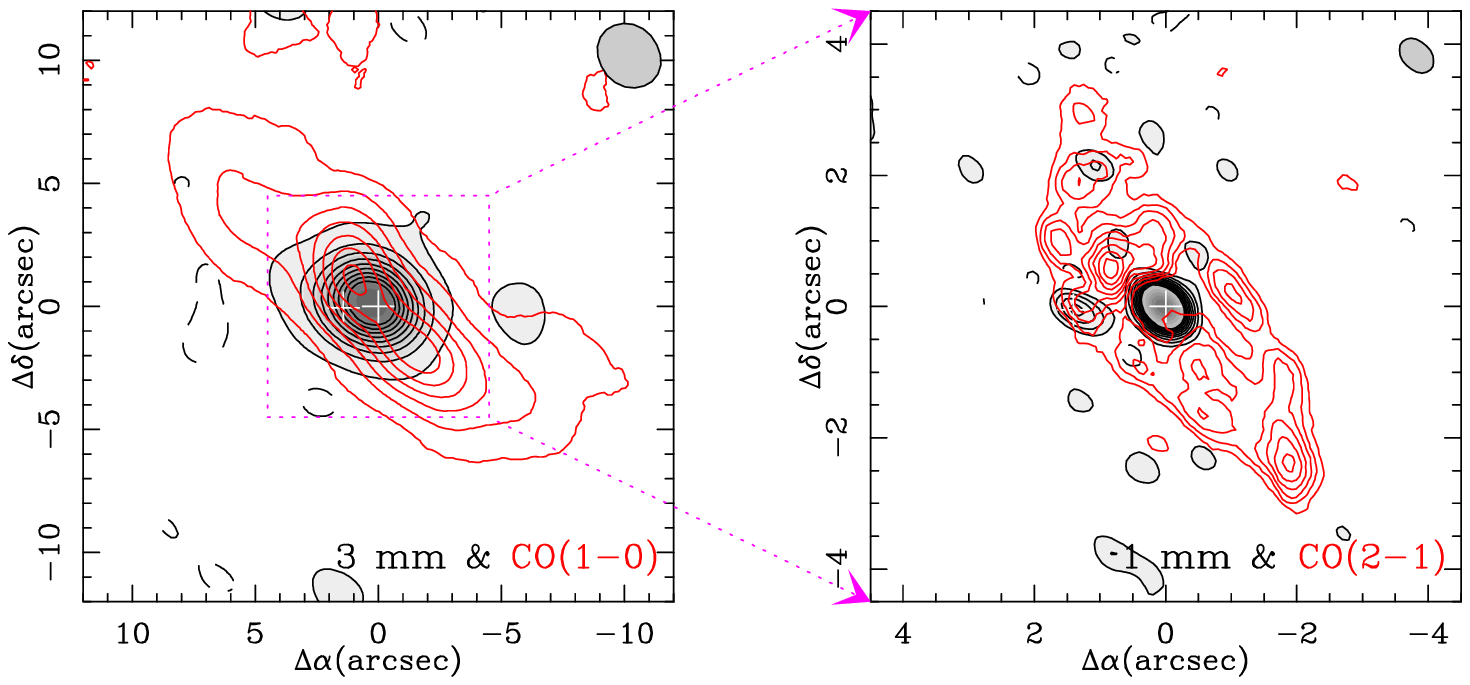}
\caption{Overlay of  the continuum (black) and CO emission (red) emission of \tcd. Crosses (+) mark the position of the AGN and jet peak emission according to the UV-FIT models (Table \ref{uvfits}). The contour levels are 0.33  to 10.23~Jy~beam\mone\ \kms, in steps of 1.65 ~Jy~beam\mone\ \kms\ for the \couc\ map; and 0.39 to 3.9~Jy~beam\mone\ \kms, in steps of 0.39~Jy~beam\mone\ \kms\ for the \codu\ map. The contour levels of the continuum are the same as in Fig. \ref{contfits}. 
\label{cont_coemis}}
\end{figure*}

\begin{figure*}
\centering
\includegraphics[width=1.6\columnwidth]{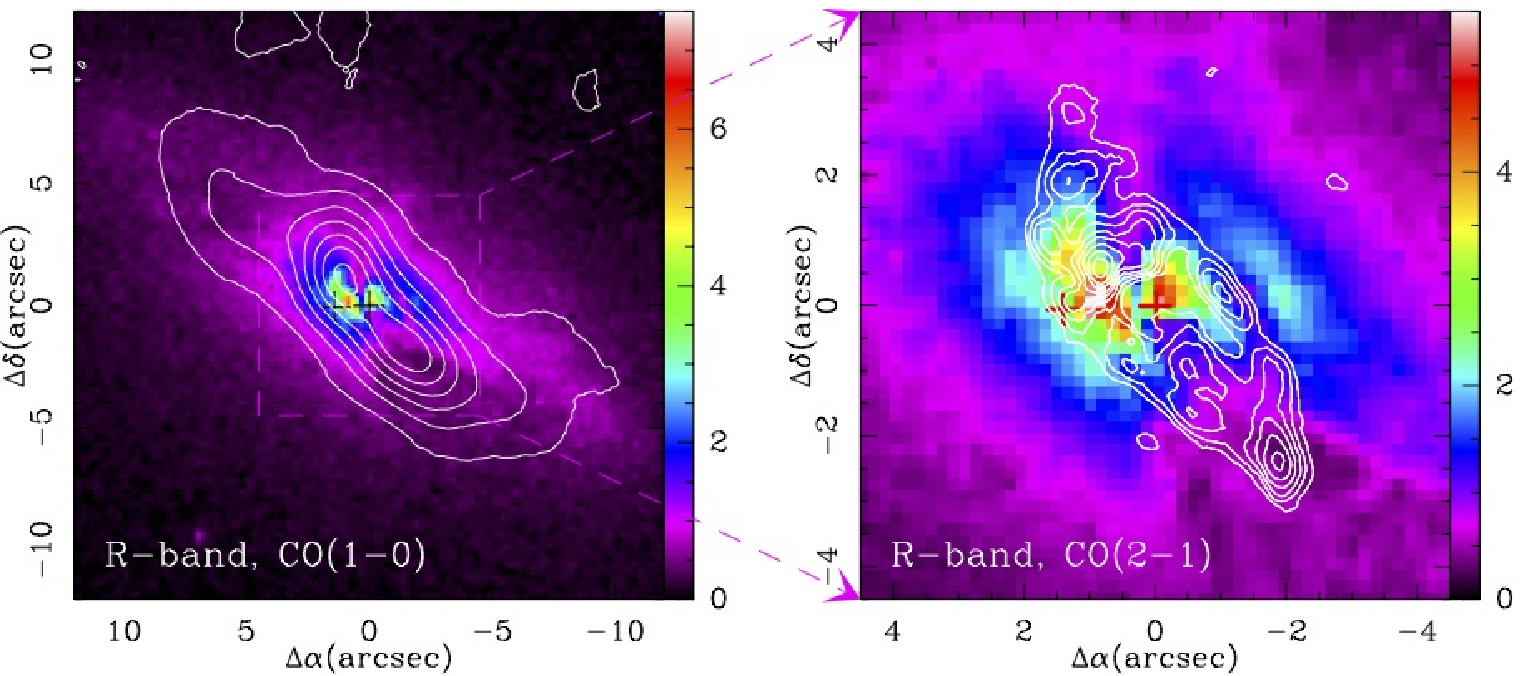} 
\includegraphics[width=1.6\columnwidth]{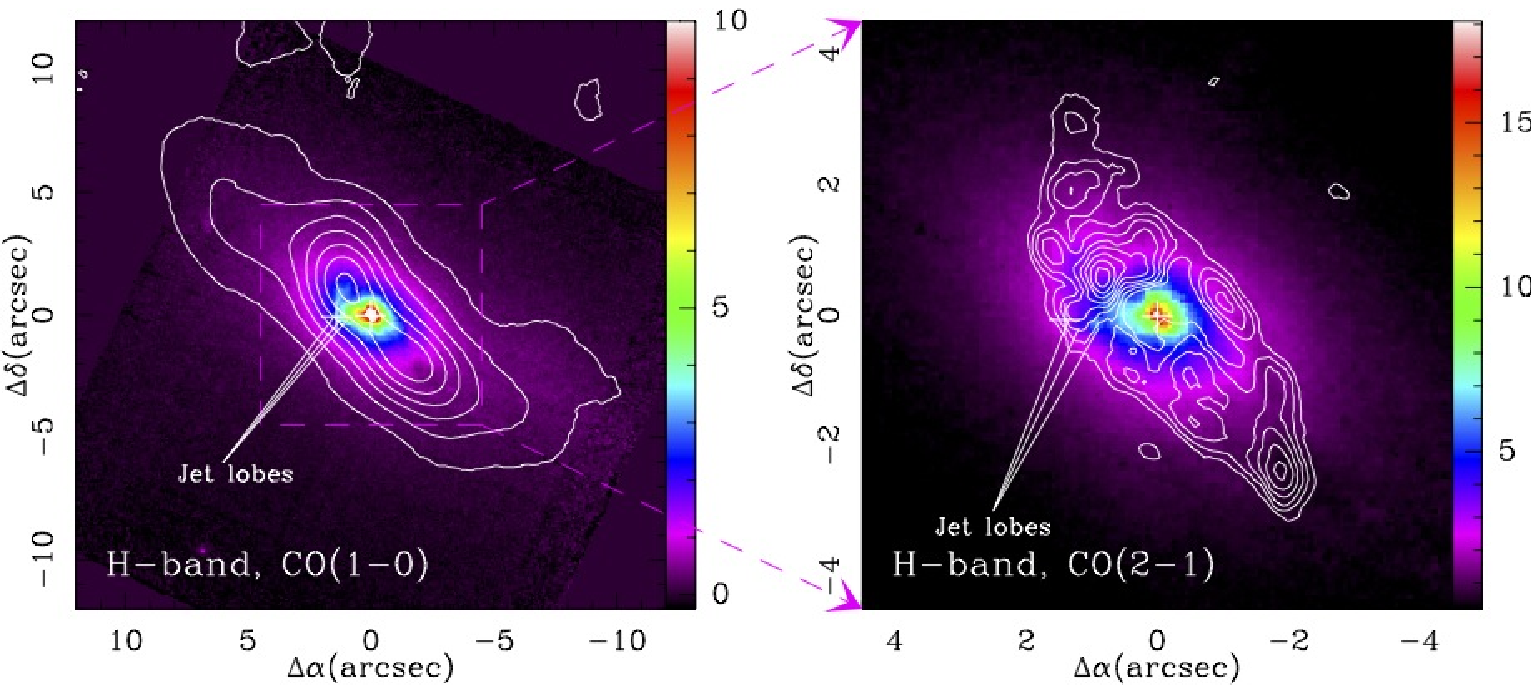} 
\includegraphics[width=1.6\columnwidth]{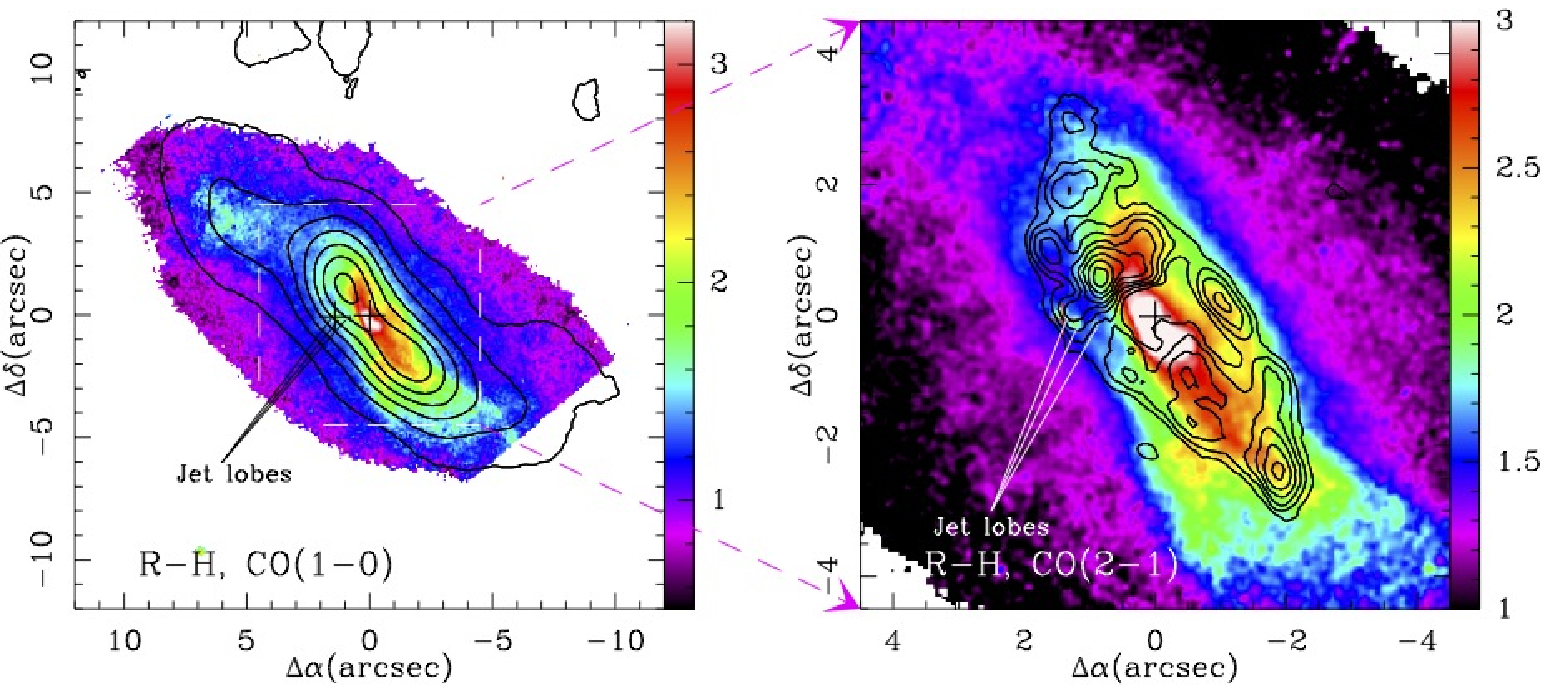} 
\caption{$R$-band \citep{Martel99}, $H$-band \citep{Floyd06}, and R--H \citep{Floyd06} images of \tcd, with the \couc\ and \codu\ emissions overlaid.  ($\Delta\alpha$,~$\Delta\delta$)-offsets in arcsec are relative to the location of the AGN. Crosses (+) mark the position of the AGN and jet peak emission according to the UV-FIT models (Table \ref{uvfits}). Solid lines mark the location of the IR counterparts of the Eastern radio lobe \citep{Floyd06}. The dashed box in the \couc\ maps mark the size of the \codu\ maps shown in the right panel. Contour levels as in Fig. \ref{cont_coemis}.  \label{rhoverlay}}
\end{figure*}

Using the fluxes of the fitted mm components and the radio emission of \tcd\ between 10 MHz and 10 GHz (available in NED), we searched for the possible origin of the mm emission of \tcd. {  The 10 MHz to 10 GHz emission of \tcd\ follows a power law with spectral index $\simeq$$-0.65$$\pm$0.1. When the total fluxes (AGN plus jet) at 1~mm and 3~mm are added to the data, the spectral index ($-0.62$$\pm$0.1) is still consistent with synchrotron emission. The 1~mm and 3~mm fluxes of the AGN and jet components, considered separately, are also consistent with synchrotron emission and with the results of \citet{Floyd06}, who showed that the emission from the jet was caused by synchrotron radiation even at NIR wavelengths.}

\section{CO maps}
\label{comaps}
\subsection{Distribution of the molecular gas}
\label{dismolgas}

We mapped the \couc\ and \codu\ emission of \tcd\ with spatial resolutions of 2.4$\arcsec \times$2.8$\arcsec$ and 0.44$\arcsec \times$0.61$\arcsec,$ respectively. 
According to our data, \tcd\ shows \couc\ and \codu\ emission above significant  ($>$3$\sigma$) levels from $v$-$v_o$=--380 to +500~\kms, and $v$-$v_o$=--280 to +450~\kms\ , respectively. Figure \ref{cointeg} shows the integrated \couc\ and \codu\ emission maps 
 within these velocity ranges. 
The \couc\ emission extends to radius$\sim$12$\arcsec$ ($\sim$10.5 kpc) from the AGN, while the \codu\ emission extends to radius $\sim$4$\arcsec$ ($\sim$3.5 kpc). 
 Even though the \couc\ emission suggests a smooth morphology, the higher resolution \codu\ map shows a highly structured CO disk, with knots of emission and absorption regions 
  near the center. 
Figure \ref{cointeg} shows that the distribution of \couc\ in \tcd\ is consistent with an elongated disk-like source, with a change in position angle ($PA$) at radius $\sim$7$\arcsec$: $PA_{\rm{morph}}$$\sim$40\grados for the inner region, $PA_{\rm{morph}}$$\sim$55\grados\ for the outer region, suggesting a slightly warped molecular gas distribution. The possible presence of a warp is studied in detail in Sect.~\ref{kinco} along with the kinematics of the CO gas. 
{   
The higher resolution \codu\ map shows that the $PA_{\rm{morph}}$ in the inner $\lesssim$3$\arcsec$ radius is 55\grados, and changes to $\sim$40\grados\ at radii $\sim$3-4$\arcsec$. This abrupt change in orientation suggests a ridge in the CO disk. Inspection of the \couc--\codu\ overlay (bottom panels of Fig. \ref{cointeg}) suggests that the size of the beam might have hidden the ridge in the \couc\ map.
 }
 
 
\citet{Evans99b} observed that the \couc\ emission in \tcd\ was produced by a 7$\arcsec$ (6 kpc) diameter disk around the core of \tcd\ \citep[see also][]{Evans05}. The large difference between their measurements and our observations is due to the better sensitivity and resolution of the PdBI \couc\ data. 
Using single-dish, JCMT spectra, \citet{Papadopoulos08} measured $\sim$3 times more \codu\ flux in \tcd\ than we see in our \codu\ map, which suggests that the difference in the extension of the {  \couc\ emission (below 3$\sigma$ at radius $\sim$12$\arcsec$) and \codu\ emission (below 3$\sigma$ at radius $\sim$4$\arcsec$)}  in our data might be due to the sensitivity and PdBI filtering of the \codu\ emission on large scales. {  With the B-configuration, we loose flux at distances $\gtrsim4\arcsec$ at 1 mm. } 

\citet{Papadopoulos10} measured extremely high ratios for the CO lines: \cosc/\cotd\ and \coct/\cotd\  in \tcd\ {  (R$_{65/32}$=1.3$\pm$0.54, R$_{43/32}$=2.3$\pm$0.91).} 
Figure \ref{cont_coemis} shows the superposition of the mm continuum and CO emission maps. The \codu\ map shows a region  in front of the 1 mm core without CO emission, suggesting a CO absorber against the core. 
 The \couc\ map shows fainter emission towards the center, consistent with an absorber, although the flux  variations are weaker than the \codu\ map because of the larger size of the \couc\ beam.
 Our data suggest that the high line ratios measured by \citet{Papadopoulos10} might be caused by the CO absorber towards the central regions of \tcd, because lines with low-$J$ values are more affected by absorption than the high-$J$ lines, which in turn yields overestimated ratios. 

Figure \ref{rhoverlay} shows the \couc\ and \codu\ maps superposed on the {\it HST} $R$-band, $H$-band, and $R$-$H$ color images of \tcd. 
The $R$-band image shows two optical absorption systems: a disk-like dust structure in the inner $\lesssim$1 kpc radius, and the large-scale ($\gtrsim$10 kpc radius) tidal features responsible for the dust lanes that criss-cross the galaxy in front of the nucleus \citep{Heckman86, Smith89, Evans99b, Floyd06}. The dust disk is oriented along the northwestern edge (where obscuration is higher) towards us. 
  The $H$-band image shows that the IR counterparts of the Eastern jet radio lobe, which are on the approaching side of the radio source, are emerging from the southeastern side of the disk \citep{Floyd06}. 
The overlays of the CO emission and the $R$, $H,$ and $R$-$H$ maps show that the \codu\ emission follows the edge of the dust disk in the northwest, suggesting that the molecular gas is associated with the dust disk. 
   The $R-H$ map shows that the extinction follows the \couc\ emission contours up to distances of at least $\sim$12$\arcsec$ (radius).  Furthermore, the extinction in \tcd\ also follows the warped structure of the CO disk, supporting that the dust and CO disk are associated up to distances $\gtrsim$12$\arcsec$ (10.5 kpc). 
Hence, our observations suggest that \tcd\ hosts a warped molecular gas (and dust) disk that extends at least  $\sim$12$\arcsec$ in radius. At the largest distances, the disk is seen highly inclined. 
 The same warped structure is seen in other gas and star formation tracers of \tcd\ (see the NUV and 8 \mum\ emission maps in Sect. \ref{secsfe}). In Sect. \ref{kinco}, we study the dynamics of the CO emission in detail to determine whether it is consistent with a rotating, warped disk.
 
\begin{figure}
\centering
\includegraphics[width=0.85\columnwidth]{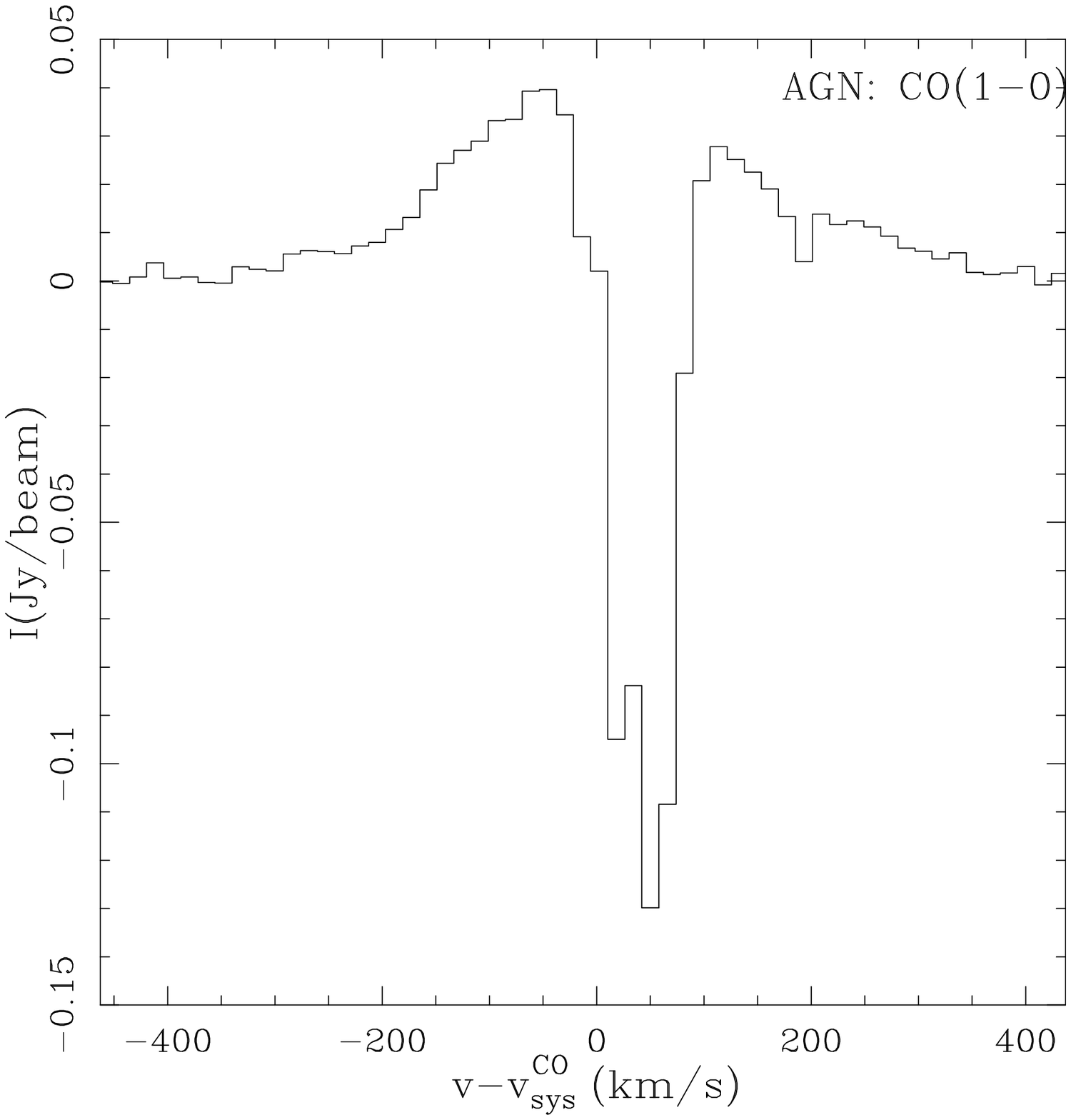} 
\includegraphics[width=0.85\columnwidth]{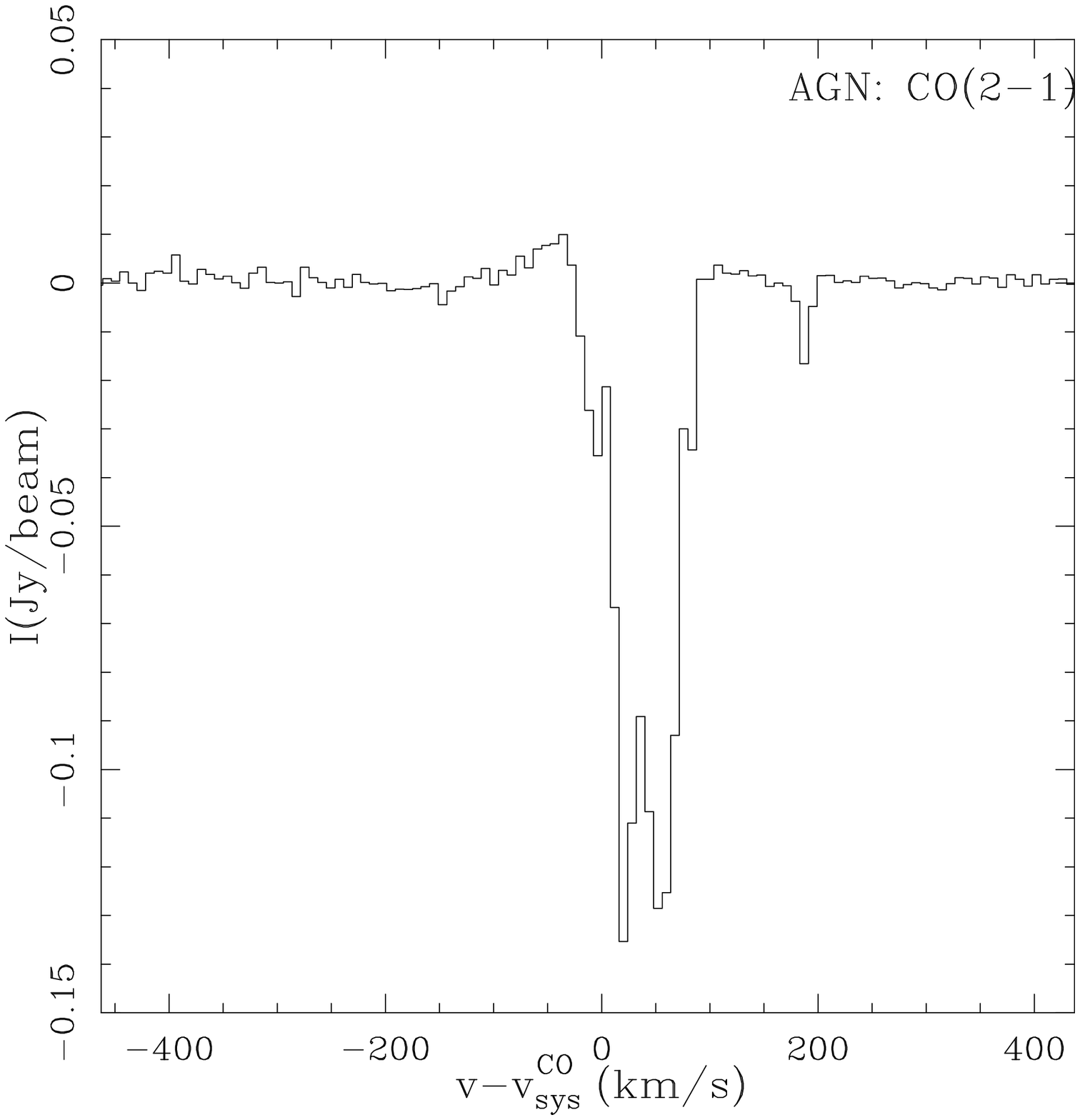}  
\caption{Spectra of the \couc\ and \codu\ lines (with the continuum subtracted) toward the AGN of \tcd. For clarity, we have zoomed into the velocity ranges where the emission is detected. 
\label{spectrum}}
\end{figure}

\subsection{Mass of molecular gas}\label{mass}

The total \couc\ flux emitted by \tcd\ is 53~Jy \kms. The equivalent \couc\ luminosity is $L$$\arcmin_{\rm{CO}}=4.8\times10^9$ K \kms\ pc\mtwo\ Ê\citep{Solomon97}. Applying the Galactic ratio of \hd-mass to CO-luminosity \citep[4.6 \msun/K \kms\ pc$^2$][]{Solomon87}, the H$_2$ mass in \tcd\ is \mhd=2.2$\times$$10^{10}$ \msun. 
  The column density of hydrogen atoms derived from CO is $N_{\rm{H,CO}}$=2.8$\times$$10^{22}$ cm$^{-2}$.  
\citet{Evans99b} measured a total \couc\ flux emitted by \tcd\ of 52$\pm$11 Jy \kms\ in a 74$\arcsec$ diameter aperture, suggesting that we are including all CO emission of \tcd\ in our observations.




 The dynamical mass ($M_{\rm{dyn}}$) inside a radius, $R$, can be derived as  $M_{\rm{dyn}}$=$R$$\times$$\varv_{\rm{rot}}^2$/G, where G is the gravitational constant, and $\varv_{\rm{rot}}$ is the de-projected radial velocity at the edge of the disk. The fits to the velocity maps of \tcd\ (Sect. \ref{kinco}) show that the circular velocity of the CO disk is 251 \kms\ at $R$=9.2$\arcsec$ (8 kpc). The inclination angle of the disk at that radius is $i$=58\grados\ (Sect. \ref{kinco}). Therefore,  $\varv_{\rm{rot}}$$\simeq$300 \kms, and
   $M_{\rm{dyn}}$$\simeq$2.0$\times10^{11}$ \msun. Removing the mass of molecular gas derived above and neglecting the contribution from dark matter, the spheroidal stellar mass within $R$=8~kpc is $M_{\rm{sph}}$$\simeq$1.8$\times10^{11}$ \msun, consistent with the total stellar mass derived from stellar population fits to the optical spectra \citep[$2.8\times 10^{11}$ \msun,][]{Tadhunter11}, and the M$_{\rm{bulge}}$ derived from optical photometry \citep{Bettoni03}. 

\subsection{Determination of the systemic velocity}
\label{detvsys}

The first measurements of the  \vs\ of \tcd\  were made by \citet{Sandage66} and \citet{Burbidge67}. Using optical emission lines, they obtained \vs=13\,500 \kms, which is also the central velocity of the deepest \hi\ absorption. That value was adopted until 2005, when \citet{Emonts05} used high-resolution long-slit spectra of the ionized gas emission lines to calculate the redshift of the narrow component of the optical emission lines in the nucleus of \tcd: $z$=0.048\,6$\pm$0.001\,2, $v_{\rm{sys}}^{\rm{Em05}}$=13\,450$\pm$35 \kms. 

Our data resolve the whole molecular gas (and dust) disk that
rotates around the AGN, which allows a new determination of the systemic velocity of \tcd. Figure \ref{spectrum} shows that the \couc\ and \codu\ spectra of \tcd\ at the AGN consist of a combination of absorption and emission lines. 
Absorption lines can be formed in clouds with noncircular trajectories that cross in front of the AGN or jet. Estimations based on absorption lines, which are sensitive to a low column density of gas, may then yield incorrect values of the \vs. The CO emission lines of \tcd, in contrast, are produced ubiquitously in the rotating disk. The inferred kinematic parameters, including \vs, are thus more representative of the majority of the molecular gas. 
 Therefore, we used the \couc\ emission line (which has higher signal-to-noise ratio than the \codu\ line) toward the AGN to measure the \vs\ of \tcd.
{  After removing the absorption line from the spectrum, we averaged the velocities where the \couc\ emission line shows emission over 3$\sigma$, weighting them by the flux in each velocity channel, and obtained \vs=13\,437$\pm$10 \kms.  A Gaussian fit of the emission line produces \vs=13\,435$\pm$10 \kms. Both values are consistent with \vsco= 13\,434$\pm$8 \kms derived from the kinematic analysis of the CO disk (see Sect. \ref{kinco}), and the optically
derived $v_{\rm{sys}}^{\rm{Em05}}$. 
 Hereafter, all velocities are given with respect to \vsco=13\,434$\pm$8 \kms.}

\subsection{Kinematics of the molecular gas}
\label{kinco}

\begin{figure*}[!]
\centering
\includegraphics[width=1.8\columnwidth]{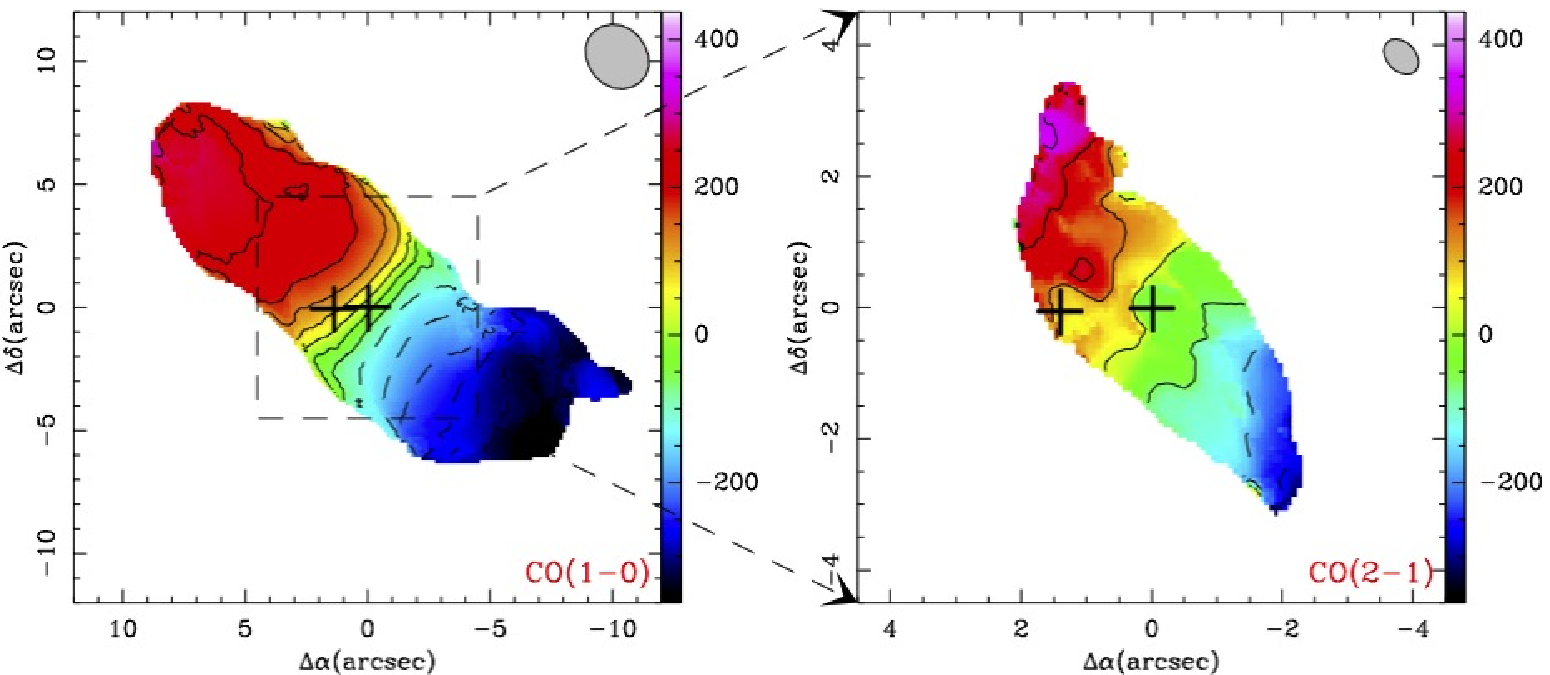} 
\caption{\couc\ (left) and \codu\ (right) velocity maps of \tcd. Velocities are given in \kms\  with respect to \vsco. Isovelocities are marked every 50 \kms\ and every 100 \kms\ for the \couc\ and \codu\ maps, respectively. 
 Crosses (+) mark the position of the AGN and jet peak emission according to the UV-FIT models (Table \ref{uvfits}). The dashed box in the \couc\ map marks the size of the\ map shown in the right panel.  \label{velmap}}
\end{figure*}

Figure \ref{velmap} shows the \couc\ and \codu\ velocity maps of \tcd.  Based on these maps, the velocity fields of the \couc\ and \codu\ emission lines are consistent with a rotating disk where the northeastern side is receding and the southwest is
approaching. {  The velocity field of the \couc\ line shows that the major axis of the disk is oriented along $PA_{\rm{vel}}$$\sim$45\grados\ in the inner 7-8$\arcsec$ (radius), and shifts toward $PA_{\rm{vel}}$$\sim$55\grados\ for larger distances, consistent with the $PA_{\rm{morph}}$ estimated from the \couc\ emission map (Sect. \ref{dismolgas}). Visual inspection of the \codu\ velocity field cannot confirm the ridge seen in the \codu\ line emission map at radius $\sim$3-4$\arcsec$. The ridge is verified by the analytic study of the CO kinematics below. 
}

\begin{figure*}
\centering
\includegraphics[width=1.85\columnwidth]{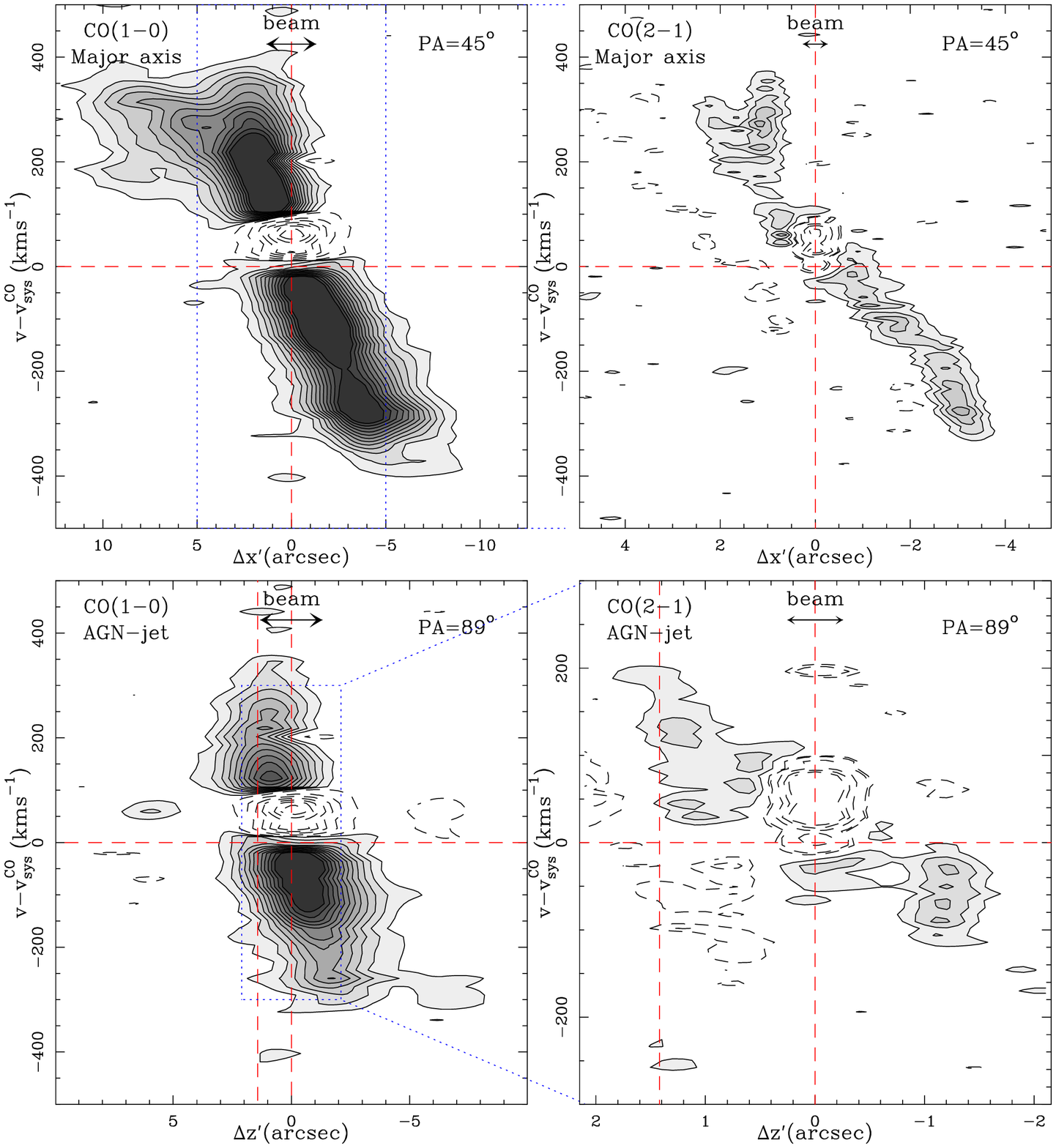}   
\caption{Position-velocity diagram of the \couc\ (left panels) and \codu\ (right panels) emission along the major axis of the molecular gas disk (top panels) and AGN-jet line (bottom panels) of \tcd. Positions ($\Delta$z') are relative to the AGN. Velocities are relative to the \vsco. The contour levels are $-3\sigma$, $3\sigma$, to $36\sigma$ in steps of $3\sigma$ for the \couc\ diagrams and $-3\sigma$, $3\sigma$ to $12\sigma$ in steps of $3\sigma$  for the \codu\ diagrams. $1\sigma=0.9$ and 1.7 mJy~beam\mone\ for the \couc\ and \codu\ diagrams, respectively.
 \label{pvdiagram}}
\end{figure*}

Figure \ref{pvdiagram} shows the position-velocity (P-V) diagrams for the \couc\ and \codu\ lines taken along the major axis of the disk ($PA$=45\grados), and along the AGN-jet line ($PA$=89\grados). The P-V diagrams along the major axis
 are consistent with a CO-emitting, spatially resolved rotating molecular gas disk, with projected peak velocities of $\pm$350\kms\ at its edges. 
 Both the \couc\ and \codu\ P-V diagrams show a spatially unresolved absorption feature centered on the AGN coordinates,  with a width of $\sim$100~\kms, as seen in the CO spectra (Fig. \ref{spectrum}).
The P-V diagram of \couc\ along the AGN-jet line shows that the AGN absorption is spread by the beam up to the jet coordinates. 
 The \codu\ P-V diagram shows two absorption features. 
  The first one is detected 
  between distances 0.4\arcsec\ and 1.6$\arcsec$ from the AGN, centered on --100 \kms, with a width of $\sim$100 \kms. 
  This feature corresponds to the blueshifted absorption detected toward the jet coordinates in the \codu\ spectrum. The second absorber is detected at the AGN coordinates, centered on velocities $\sim$200 \kms, corresponding to the narrow absorption feature seen in the \codu\ spectrum of the AGN (Fig. \ref{spectrum}).

To analyze the kinematics of CO in \tcd\ in more detail, we used the IDL tool \kin\ \citep{Krajnovic06}. 
{\it Kinemetry} divides the 2D line-of-sight velocity map of a galaxy into a series of ellipses. Each of these ellipses has a velocity profile that is decomposed into a finite series of harmonic Fourier terms:

\begin{equation}
\quad V = c_0(r) + \sum_{n=1}^{N} [c_n(r) \cos(n \psi) + s_n(r) \sin(n \psi)], 
\end{equation}

where $r$ is the radius of the ellipse, $\psi$ the eccentric anomaly, and $c_0$ the \vs. \citet{Krajnovic06} showed that to describe the velocity map, only terms up to N=3 are needed. The circular (\vc) and noncircular (\vnc) velocities can be obtained using \citep{Schoenmakers97}

\begin{equation}
\quad  V_{\rm{circ}}=c_1  \quad , \quad  V_{\rm{nc}}=\sqrt[]{s_1^2+s_2^2+c_2^2+s_3^2+c_3^2}
.\end{equation}

To find the ellipse along which the velocity field should be extracted (the best-sampled ellipse), \kin\ minimizes \vnc$^2$ \citep[$\chi^2=s_1^2+s_2^2+c_2^2+s_3^2+c_3^2$; see also][]{Carter78, Kent83, Jedrzejewski87}. The best-sampled ellipse for a given radius is then given in terms of central coordinates, ellipticity, and PA.

We set the ellipse radii in steps of 0.2$\arcsec$ for the \codu\ velocity field. For the \couc\ field we used ellipses every 0.5$\arcsec$ in the inner regions ($<$4$\arcsec$) and 1$\arcsec$ for the outer regions ($>$4$\arcsec$). The minimum radius was set to half the FWHM of the beam for both maps, and the centers of the ellipses were fixed at the AGN coordinates,{  which is implicitly assumed to be the dynamical center}. For ellipses where the {\it kinemetry} fit did not converge, we allowed variations in the radius of $\lesssim$0.3$\arcsec$ until the fit converged. All the Fourier coefficients, including \vc, where left free to vary.

\begin{figure*}
\centering
\includegraphics[width=\columnwidth]{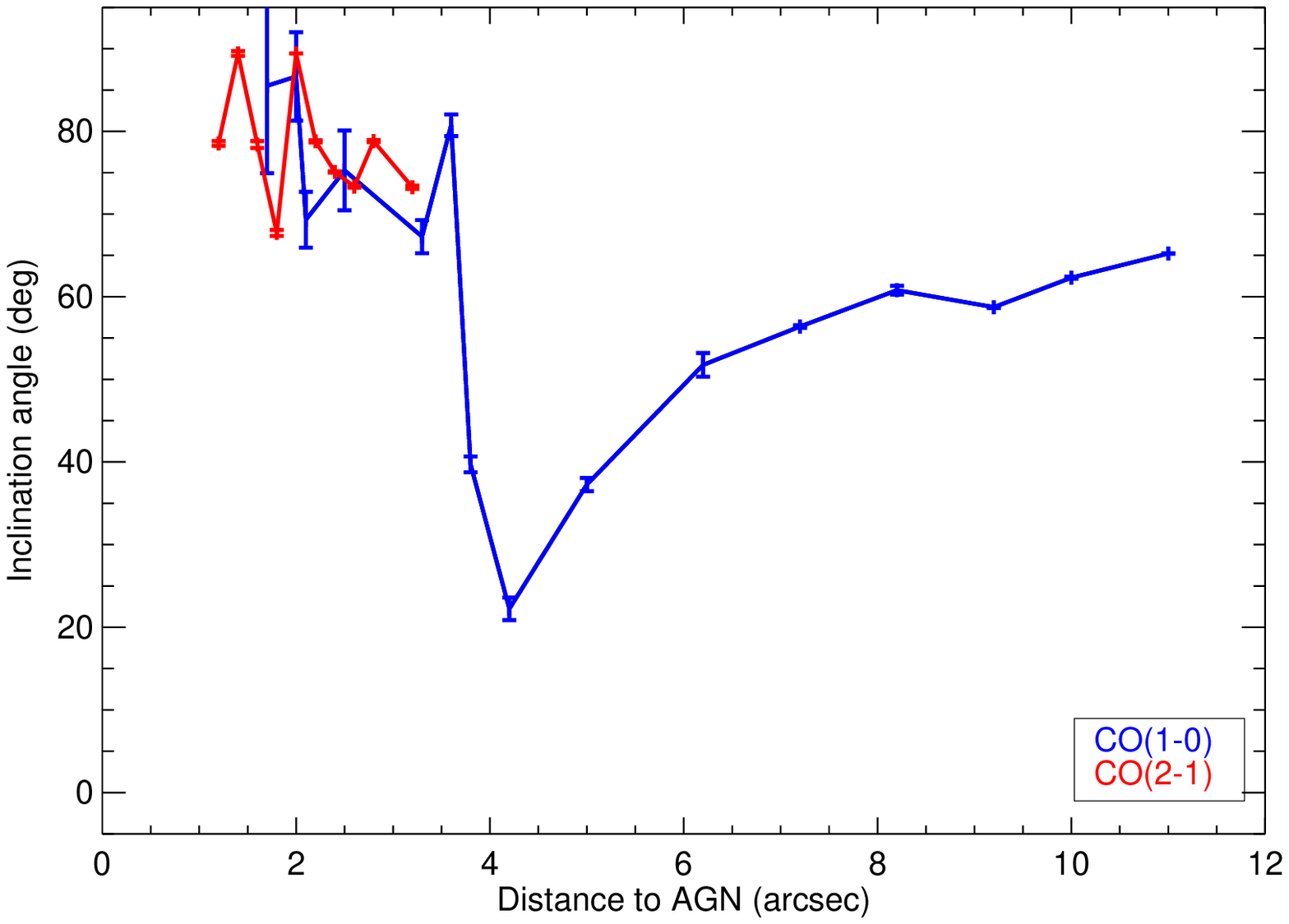} \includegraphics[width=\columnwidth]{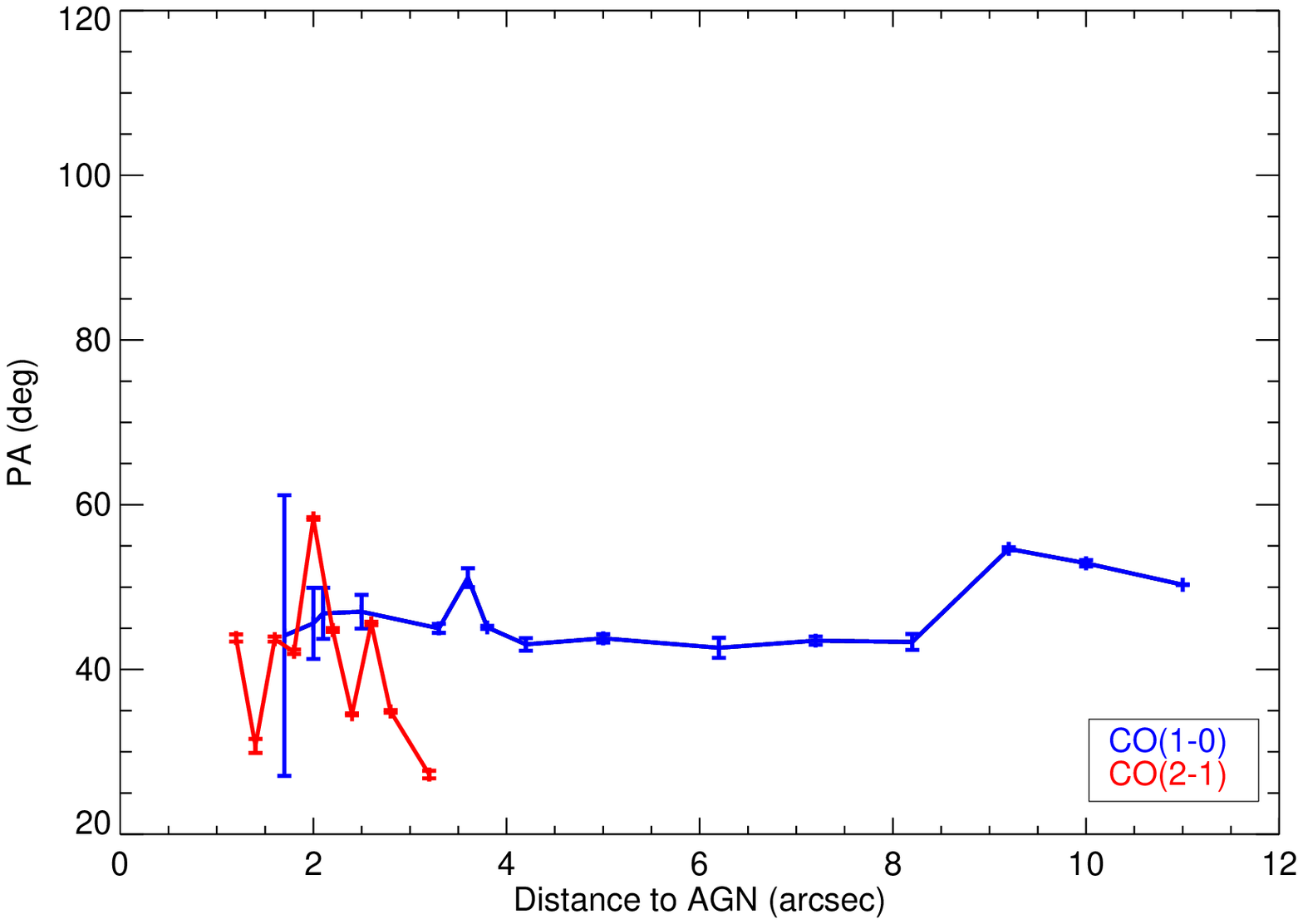}
 \includegraphics[width=\columnwidth]{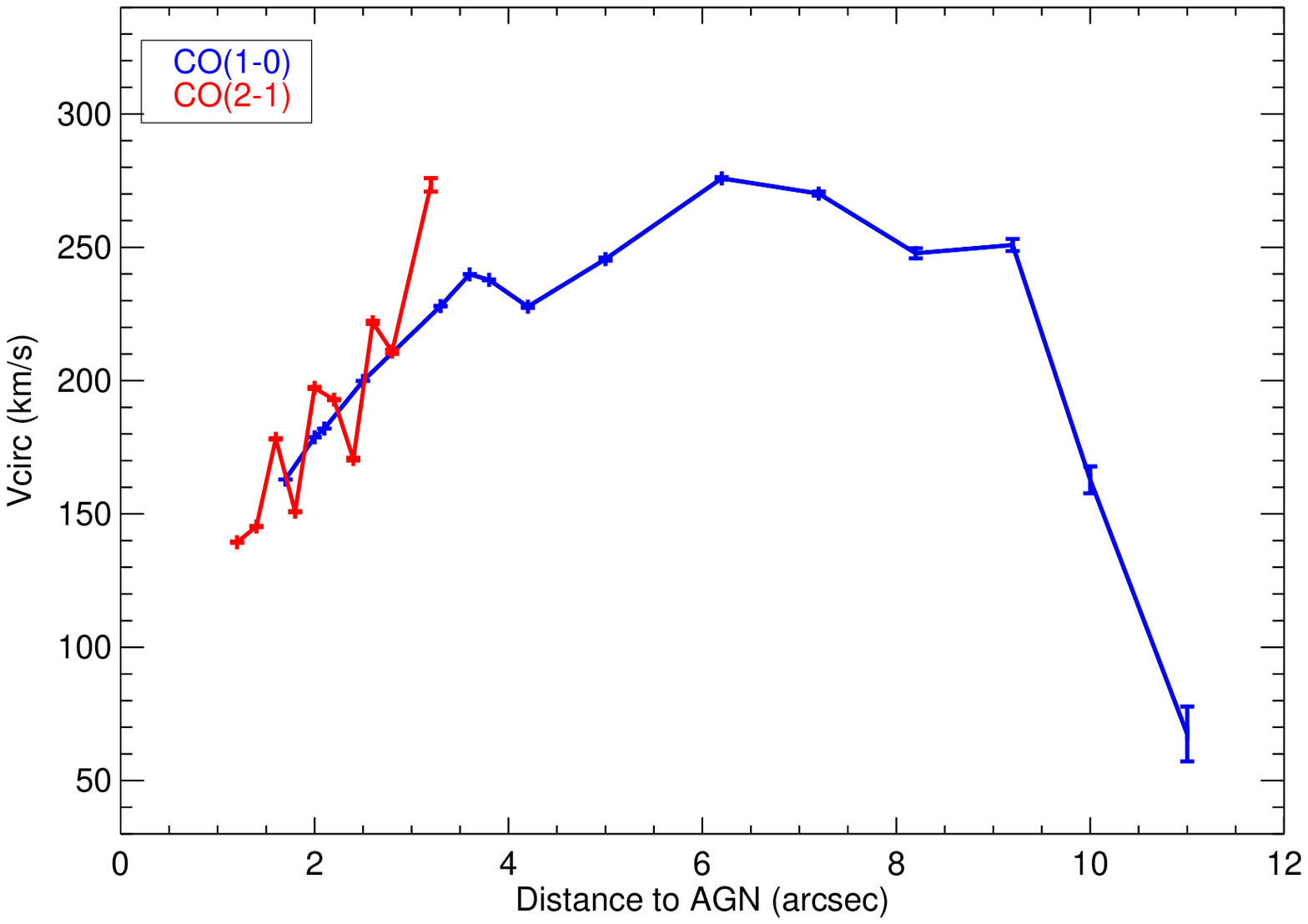}  \includegraphics[width=\columnwidth]{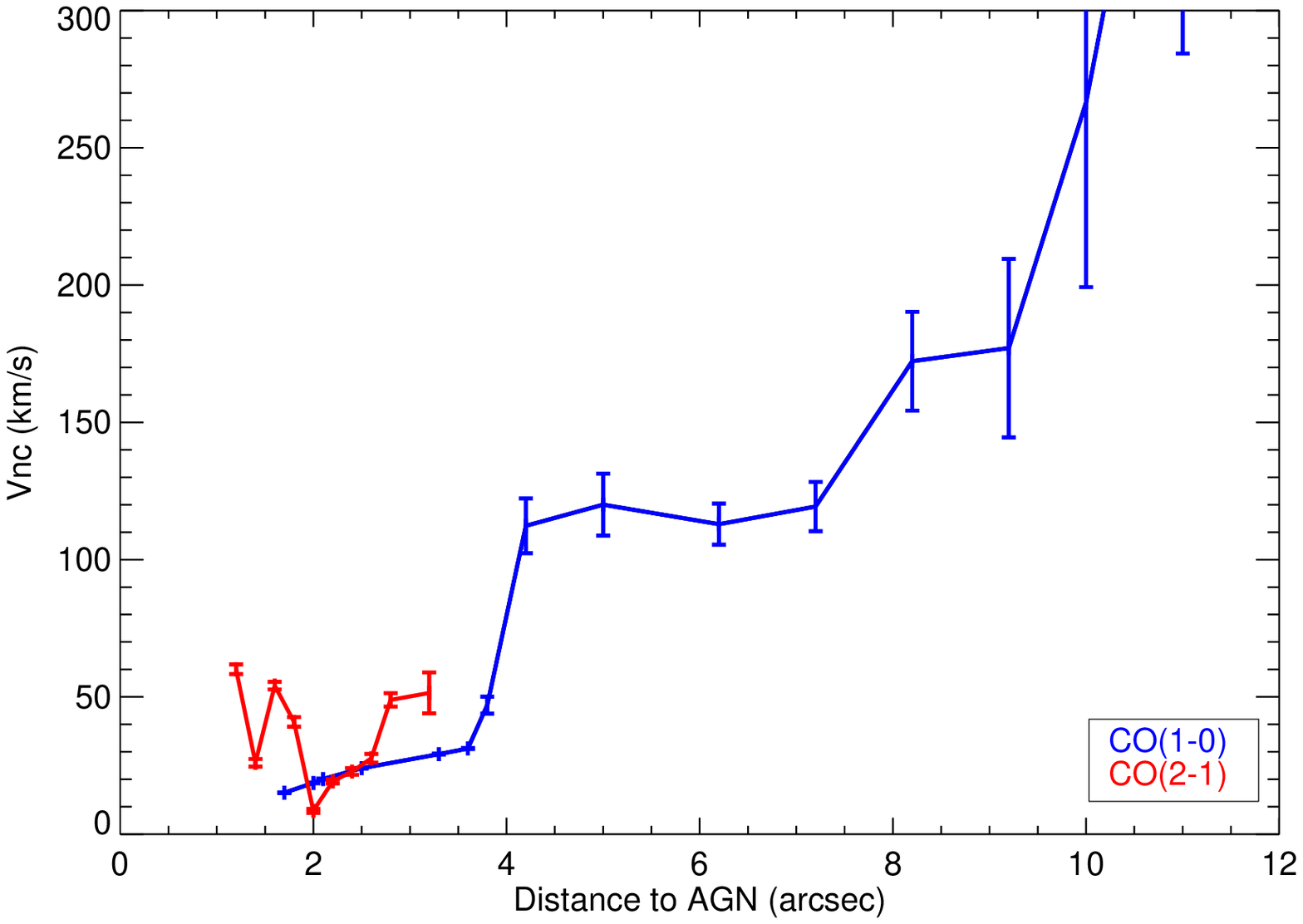}
\caption{Results of the fits with {\it kinemetry} for the inclination and position angles (top panels), the circular velocity (bottom-left panel), and the noncircular velocity (bottom-right panel).   \label{kinfig}}
\end{figure*}

Figure \ref{kinfig} shows the {\it kinemetry} fits for the inclination and position angles, and the circular and noncircular velocities for \tcd. 
The inclination angle plot shows two orientations: a roughly constant $i$$\simeq$75\grados\  for radii below 3.8$\arcsec$, and an inclination angle increasing from 20\grados\ to 65\grados\ from a radius 4$\arcsec$ to 11$\arcsec$, suggesting a twist in the CO disk. Thus, our analysis confirms the orientation proposed by \citet{Floyd06}: the southeastern side of the disk is facing us, and the eastern jet of the compact radio source is approaching. The steep change in the inclination angle along radii 4\arcsec\ and 8\arcsec\ suggests that the CO disk is warped.

The position angle is roughly constant (PA$\simeq$45\grados) for the \couc\ velocity field up to a radius of 8$\arcsec$. Beyond this radius, the diagram clearly shows a change in PA ($\simeq55$\grados), consistent with a warped CO disk. The PA fit of the \codu\ velocity map shows strong variations at small radii. These variations are likely due to small-scale structure of the velocity field and the presence of the central absorber, and not a global property of the disk. Between radii of 1$\arcsec$ and 2.5$\arcsec$, the PA of the \codu\  velocity field is consistent with the PA seen for \couc. {  Beyond 2.5$\arcsec$, the PA is $\sim$25\grados, consistent with the ridge seen in the \codu\ line emission map (Sect. \ref{dismolgas} and Fig. \ref{cointeg}). Hence, the fits of the inclination and position angles of \tcd\ support the scenario
of a warped, corrugated CO disk.}
 
 The \vc\ increase with distance to the AGN, up to 240 \kms\ at a radius of $\sim$4$\arcsec$. Beyond this radius we see a small dip in the curve followed by an increase with the distance up to 250 \kms\ at $\sim$6-7$\arcsec$. Then, the circular velocity  decreases with distance beyond 9$\arcsec$. The noncircular velocities (\vnc) constantly increase with distance all along the CO velocity field.  {  \vnc\ shows a break point at radius 4\arcsec. }
  All changes in circular and noncircular velocities occur at radii where the inclination and position angles suggest a distortion in the disk, consistent with different stages of gas settling at each orientation of the disk. 
 
{  The {\it kinemetry} fits give a total velocity ($\sqrt[]{V_{\rm{circ}}^2+V_{\rm{nc}}^2}$) of 307$\pm$33~\kms\ at R=9$\arcsec$ (8.4~kpc). 
 This corresponds to a de-projected velocity V=361$\pm$40 \kms, 
  compatible with the velocities observed in early-type galaxies \citep[full width at zero intensity, FWZI$\lesssim$800 km s\mone; e.g.,][and references therein]{Krips10,Crocker12}. 
As discussed above, \kin\ also fits the \vs\ for each ellipse of the velocity map. Based on the PA results of the \couc\ velocity map, we averaged the \vs\ fitted for each ellipse up to 8$\arcsec$. We obtained  \vsco=13\,434$\pm$8 \kms, which is consistent with the \vs\ derived in Sect. \ref{detvsys}. 
}

\begin{figure*}
\centering
\includegraphics[width=\columnwidth]{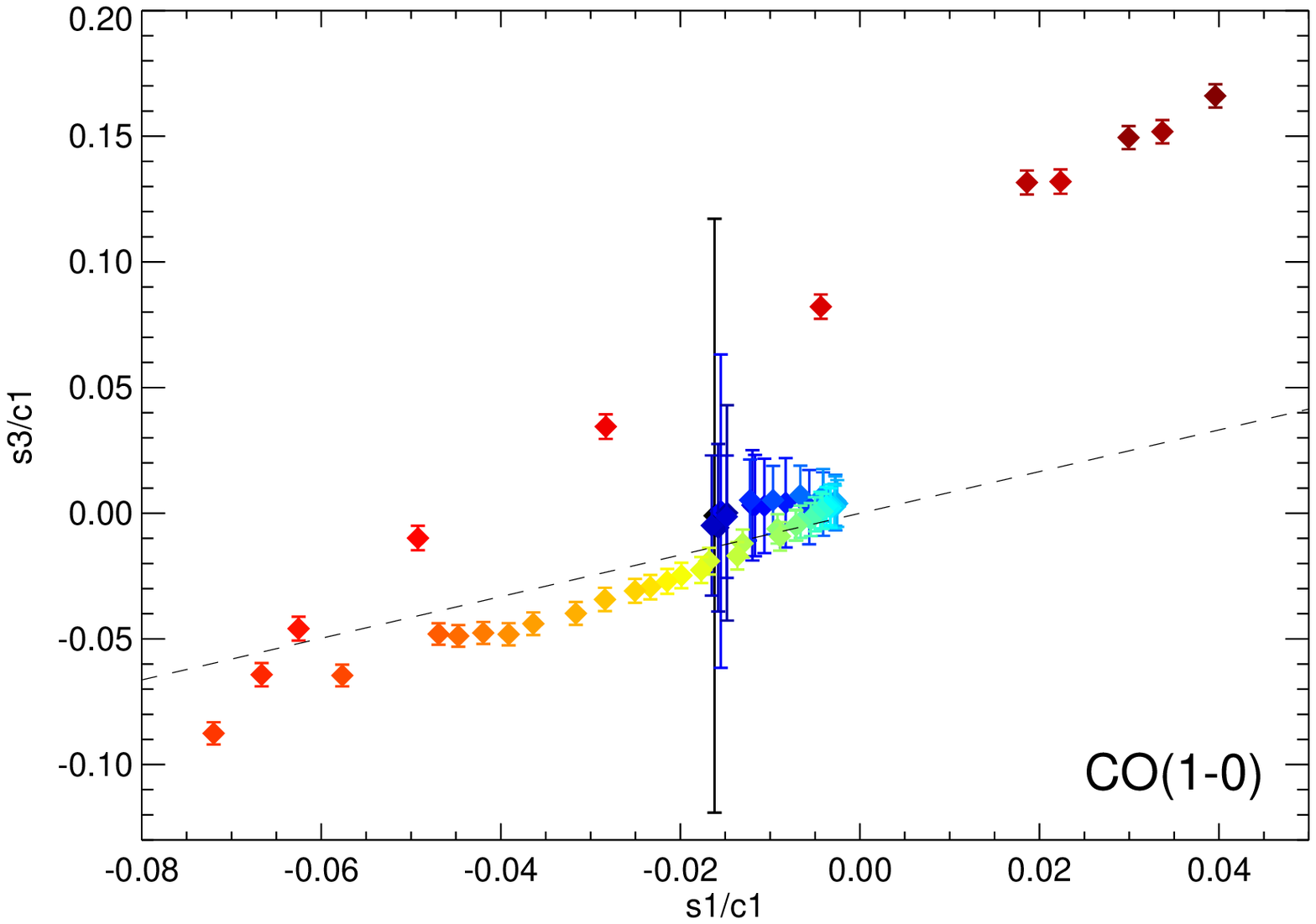} \includegraphics[width=\columnwidth]{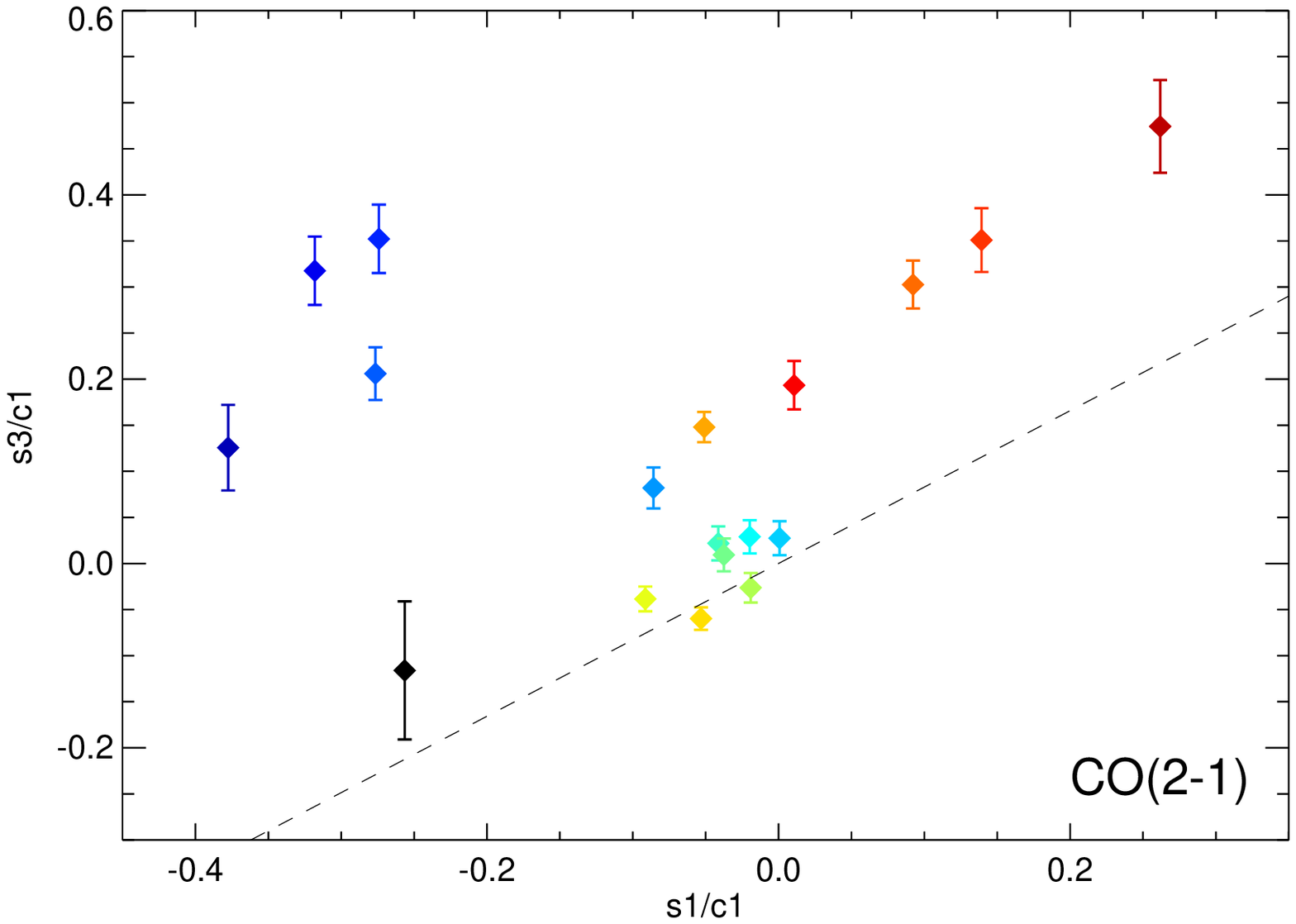}
\caption{Comparison of the s3/c1 and s1/c1 ratios of the Fourier coefficients from the {\it kinemetry} fits. 
 Radii increase from blue to red. The distance between two adjacent radii is 0.2$\arcsec$. The dashed line shows the slope of the correlation expected for a warp \citep{Wong04}.   \label{wong}}
\end{figure*}

The comparison of the first- and third-order sine coefficients of the Fourier decomposition ($s_1$, $s_3$) gives insights into the noncircular motions of the disk \citep{Schoenmakers97}. \citet{Wong04} showed that for a pure warp model, the $s_1$, $s_3$ data points follow
\begin{equation}
\hfill \frac{\delta s_1}{\delta s_3} = \frac{3q^2+1}{1-q^2} \hfill
,\end{equation}
with $q=cos(i)$. Using the method by \citet{Wong04}, we aimed to confirm the warped morphology in \tcd\ by 
 searching the kinematics of the gas for deviations from the disk rotation. We ran {\it kinemetry} a second time, fixing the inclination and position angle to their average values at the inner radii (R$<$6$\arcsec$) obtained in the first {\it kinemetry} run (77\grados\ and 45\grados, respectively). Figure \ref{wong} shows the s1/c1 and s3/c1 values obtained from this second {\it kinemetry} run. If there is a warp, s1/c1 and s3/c1 should follow a correlation with slope  $\delta s_3/\delta s_1$=0.83. This slope is illustrated in Fig. \ref{wong} with a black dashed line. The \codu\ data clearly have the warp-line slope for radii R$\geq$2.8$\arcsec$. For the \couc\ line, this behavior is seen between radii 4$\arcsec$ and 10$\arcsec$ and beyond R$\geq$ 10.8$\arcsec$, {  consistent with} the scenario of a large-scale warp in the CO disk of \tcd.


\section{Is there a molecular gas outflow in \tcd?}
\label{coldmg}

The radio and optical spectra of \tcd\ show broad ($\sim$1000~km~s\mone) components in the \hi\ absorption and ionized gas emission lines generated by fast gas outflows near ($\lesssim$1~kpc) the AGN. These outflows seem to be produced by the interaction between the radio jets of the inner radio source and the ISM of the host galaxy \citep{Morganti03, Emonts05,Mahony13}. 

\begin{figure*}
\centering
\includegraphics[width=0.85\columnwidth]{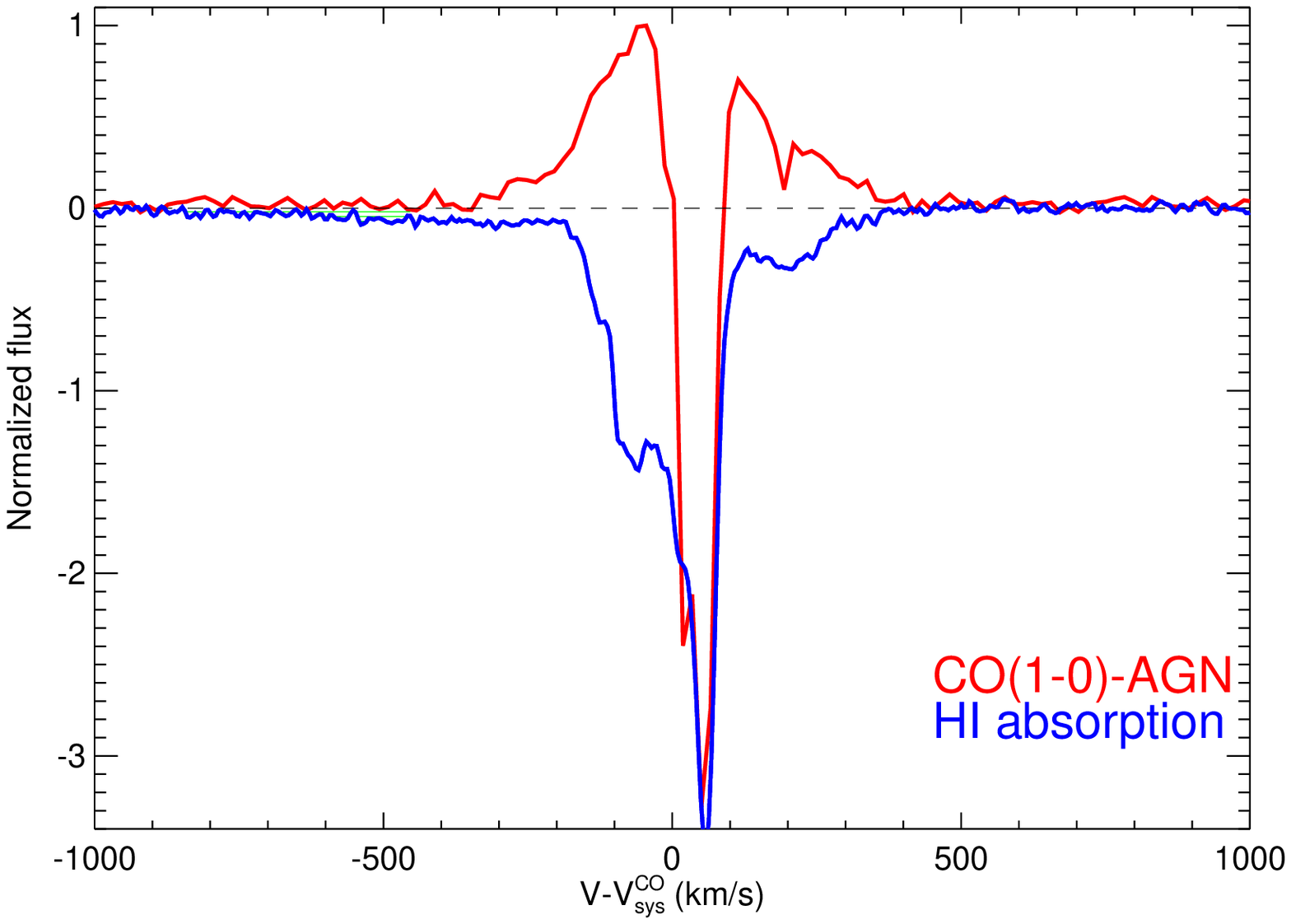} 
\includegraphics[width=0.85\columnwidth]{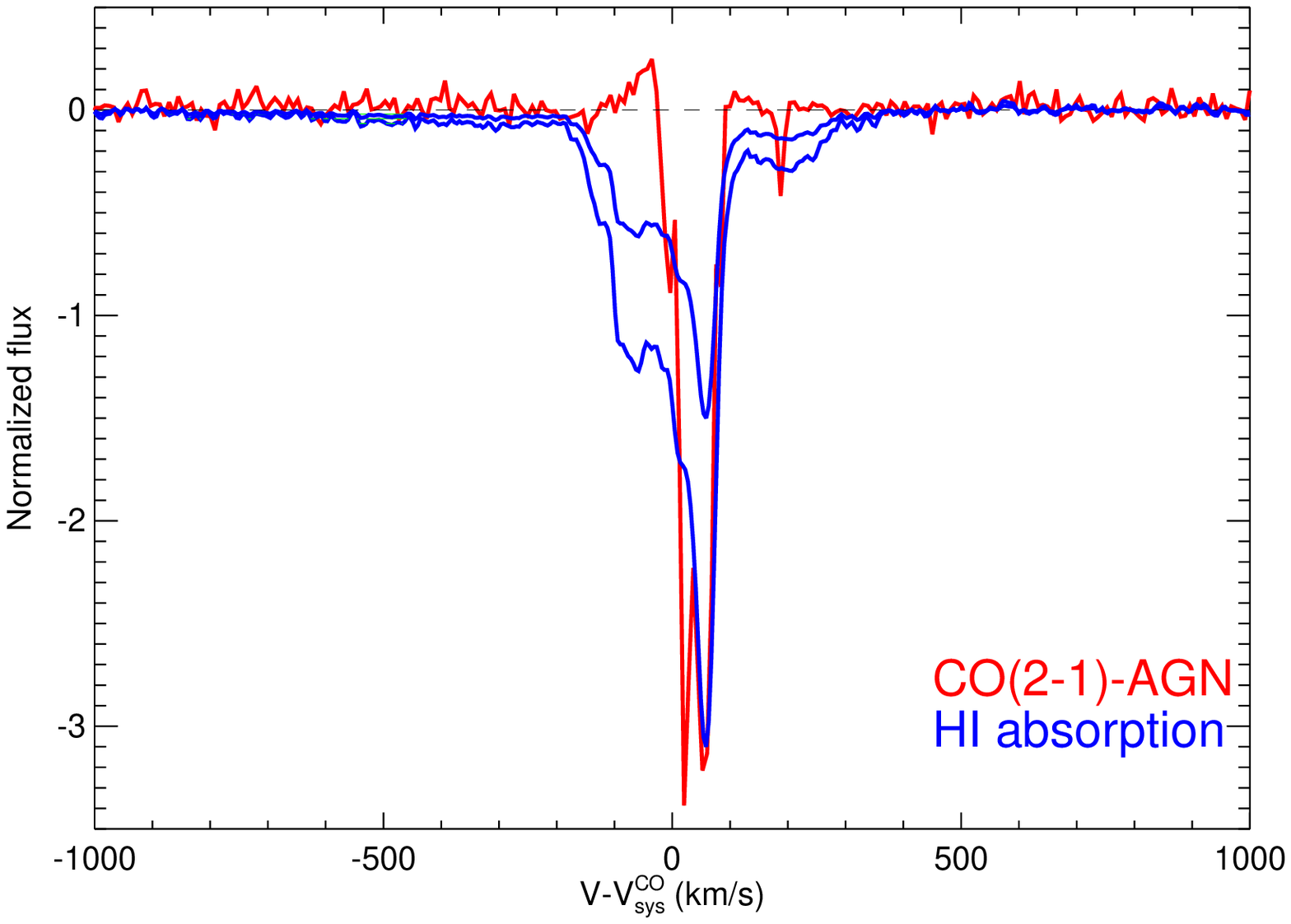} 
\includegraphics[width=0.85\columnwidth]{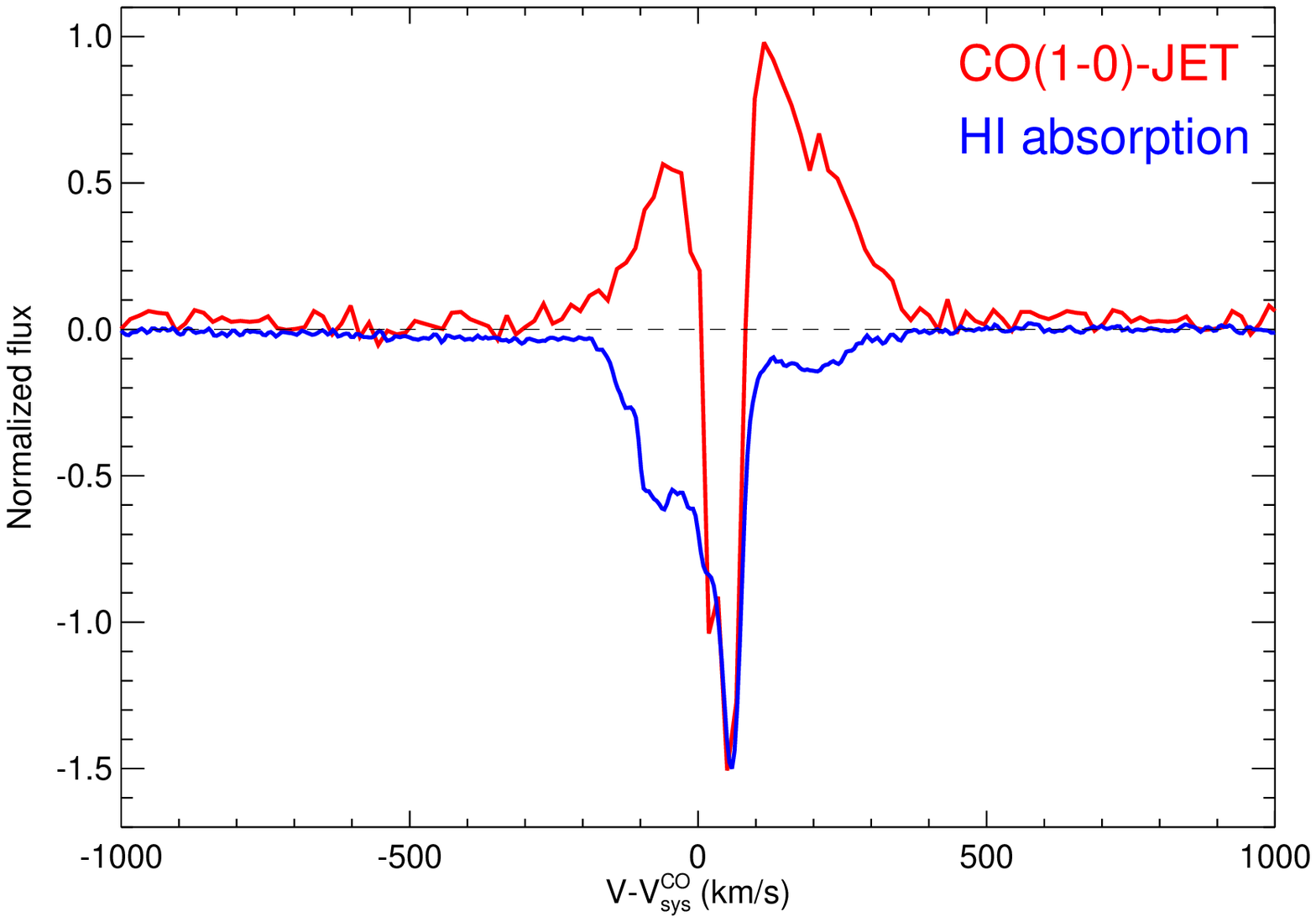} 
\includegraphics[width=0.85\columnwidth]{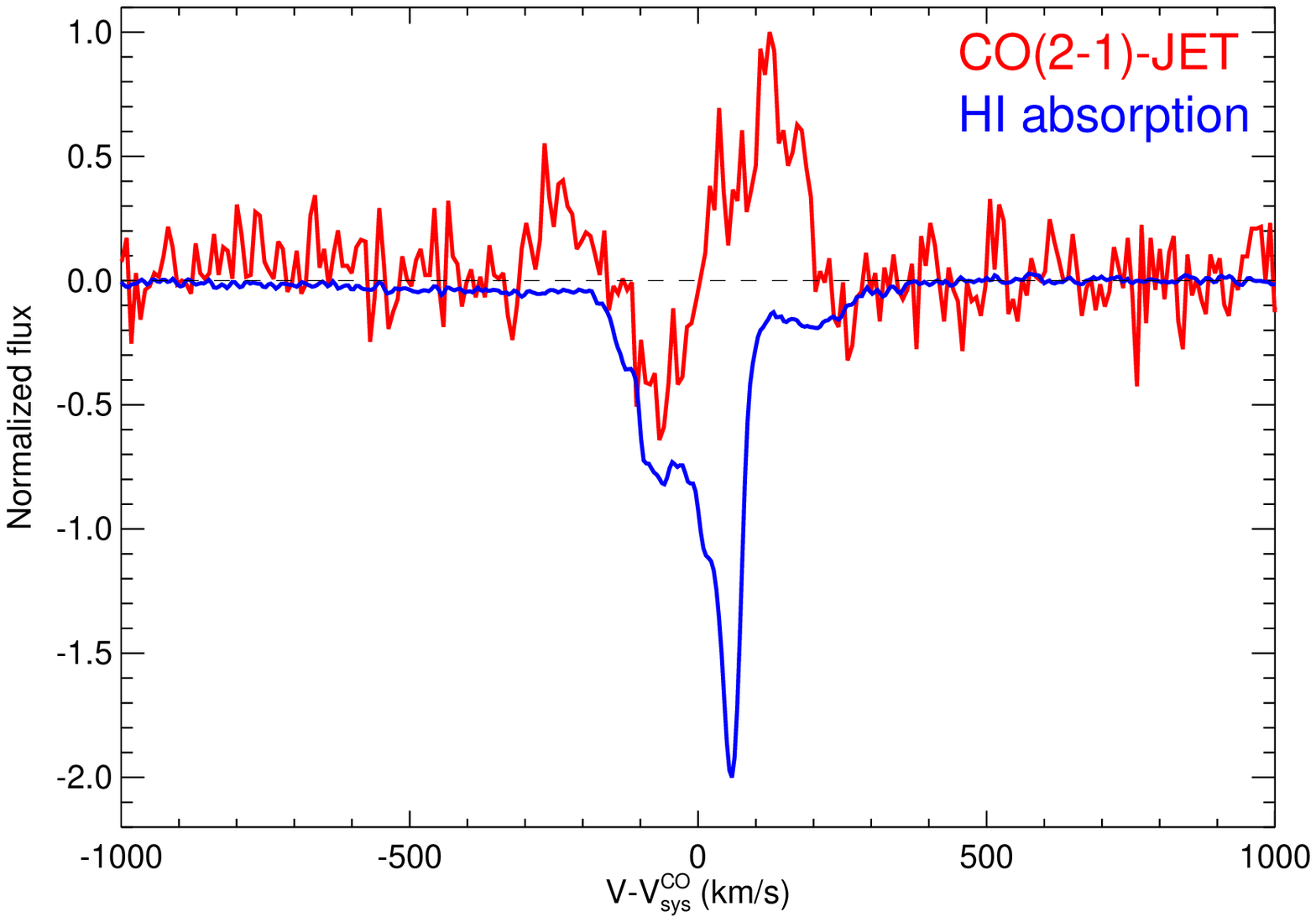} 
\includegraphics[width=0.85\columnwidth]{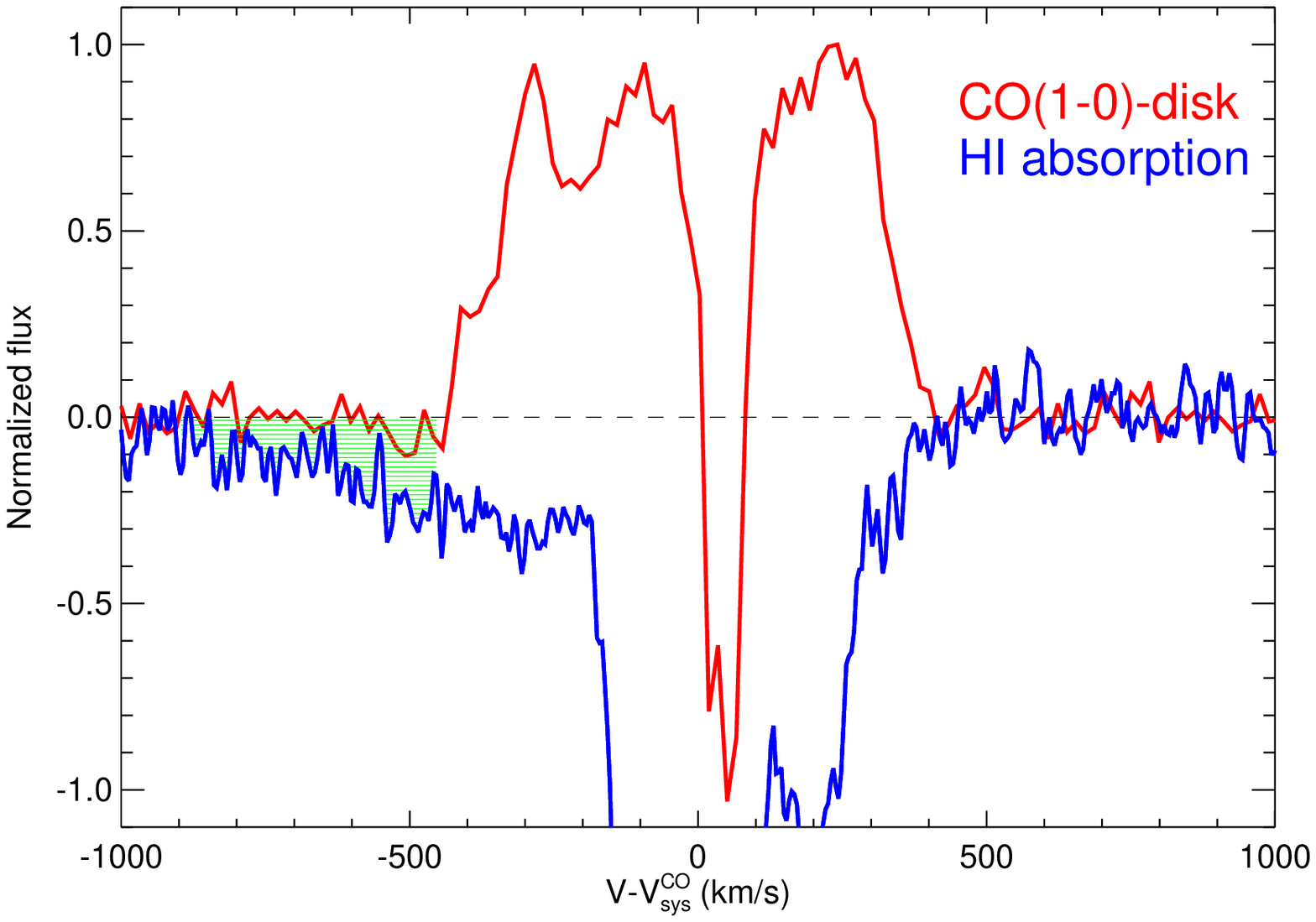} 
\includegraphics[width=0.85\columnwidth]{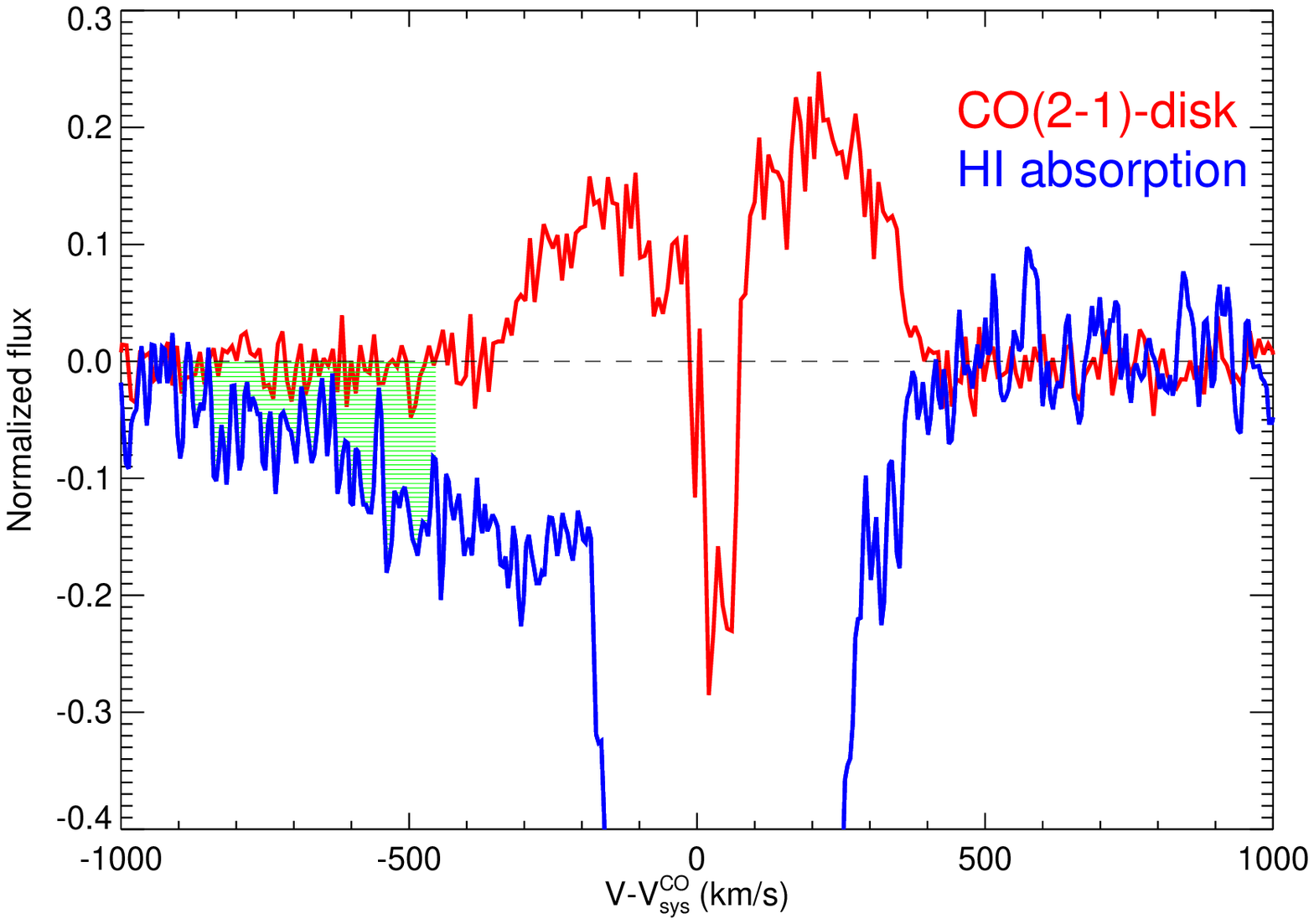}  
\caption{Overlays of the {  spatially unresolved \hi\ \citep{Morganti03} spectrum and the \couc\ and \codu\ spectra (with the continuum subtracted) toward the AGN, eastern jet, and disk regions of \tcd}. The CO spectra correspond to the regions of the AGN (top panels), jet (2nd row), and \couc\ and \codu\ emission disks (3rd row). 
The green-shaded area marks the area of the \hi\ absorption that cannot be attributed to rotation around the AGN. Fluxes are in arbitrary units. Velocities are given with respect to \vsco.
\label{allspectrum}}
\end{figure*}

{  Figure \ref{allspectrum} shows the overlay of  the \couc\ and \codu\ spectra toward the AGN eastern jet and CO disk regions, with the spatially unresolved \hi\ spectrum \citep{Morganti03}.} 
All the CO spectra show emission consistent with a  FWHM of $\sim$300-350 \kms\ Gaussian, combined with a strong, structured absorption feature. 
 The central \couc\ and \codu\ absorptions at the AGN and the \couc\ absorption at the jet show a similar structure, with central velocities $v$-\vsco$\sim$+40~\kms, and a FWHM of $\sim$60 km~s\mone. This absorption feature consists of two components separated by $\sim$40\kms, with FWHM$\sim$20-30~\kms. The AGN spectra for the \couc\ and \codu\ lines also show a narrow (FWHM$\simeq$30\kms) absorption line at $v$-\vsco$\sim$+200~\kms. 
 The \codu\ absorption at the coordinates of the jet  has a FWHM
of $\simeq$80\kms, and it is $\sim$10 times fainter and blueshifted by $\sim$100 \kms\ with respect to the central absorptions in the rest of the spectra.


\begin{figure}
\centering
\includegraphics[width=0.85\columnwidth]{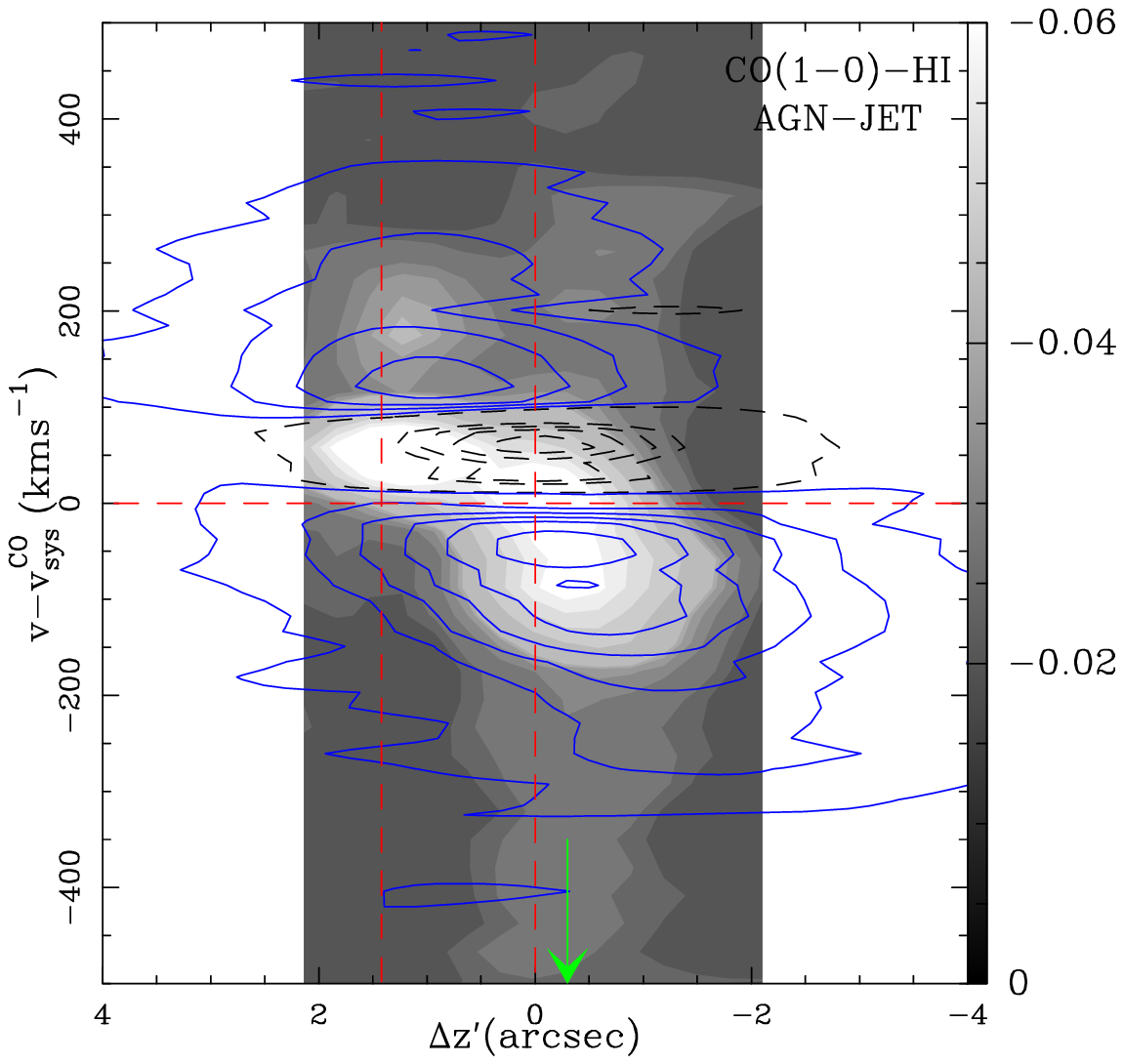} 
\caption{{Overlay of the \couc\  (contours) and the \hi\ \citep[Fig. 1 from][gray scale]{Mahony13} P-V diagrams along the AGN-jet line (PA=89\grados). Positions ($\Delta$z') are relative to the AGN. Velocities are relative to the \vsco. The vertical red lines mark the coordinates of the AGN and jet detected at mm-wavelengths. The \hi\ outflow component, seen against the western jet, reaches  --900 \kms. Its position is marked by the green arrow on the lower X axis. }
\label{pvcohi}}
\end{figure}

  The \hi\ spectrum of \tcd\ shows absorption features against the jets of the compact radio source, which seem to be produced by absorbers in the dust disk \citep{Baan81, Haschick85, Beswick02, Beswick04, Floyd06}. 
Figure \ref{pvcohi} shows the overlay of the \couc\ and \hi\ P-V diagrams along the AGN-jet line (PA=89\grados), and the location of the \hi\ outflow. 
The \hi\ absorption toward the core is centered on $v$-\vsco$\sim$+50~\kms\  and has FWHM$\sim$80~\kms\ , consistent with the CO absorption at the same coordinates. The \couc\ beam covers all of the eastern jet radio lobe \citep[regions 1 through 6 in][see also the IR jet lobes in Fig. \ref{rhoverlay}]{Beswick04}. At these coordinates, the \hi\ spectrum shows absorptions with centers ranging from $v$-\vsco=$-100$~\kms\ to 60~\kms, and FWHM from 20 to 50~\kms. 
 All the CO absorption features seen toward the jet and core of \tcd\ are associated with the \hi\ absorption components that
originated in the disk. The CO spectra show no indications of outflowing molecular gas either in the absorption or emission lines. 

 As discussed in the previous section, the kinematics of CO in \tcd\ is consistent with a large, corrugated, warped disk with circular rotation around the core. The bottom panels of Fig. \ref{allspectrum} show that $\sim$400~\kms\ of the \hi\ outflow component cannot be explained by rotation of the disk. 
None of the CO lines show traces of outflowing molecular gas in \tcd\ either in absorption or emission at the velocities of the \hi\ outflow anywhere in the galaxy (see also Fig. \ref{pvcohi}). 
{  It should be noted that our data do not have the spatial resolution needed to separate the spectra of the AGN and the western jet lobe, where the \hi\ outflow has been detected. Therefore we used our AGN spectrum to establish a limit on the mass of the CO outflow in \tcd.}
Based on the \couc\ spectrum, the 3$\sigma$ upper limit to the molecular gas mass of a 400 \kms\ cold-\hd\ emission outflow is $\lesssim$7.1$\times10^8$ \msun, 
that is, 3.2\%\ of the total of cold-\hd\ mass. 
 For comparison, the mass ratio of the outflow detected in \object{Mrk~231} is 6\%\ \citep{Feruglio10, Cicone12}, almost twice as high as the 3-$\sigma$ upper limit for \tcd; the ratio measured in \object{NGC~1266} is 2.2\% \citep{Alatalo11}, and the ratio for \object{IC~5063} is somewhere between 4.5\%--26\% \citep{Morganti13}. 
\citet{Mahony13} estimated that the mass outflow rate of the \hi\ outflow is $\dot{M}$=8--50 \msyr, assuming velocities between 100 and 600 \kms\ for the gas. If we compare the CO and \hi\ spectra in the region where the \hi\ outflow was detected (Fig. \ref{pvcohi}), the \hi\ outflow velocities beyond the rotation of the disk range from $\sim$--300 to $\sim$--900~\kms, with respect to \vsco\ (i.e. $v_{\rm{outflow}}$=0--600 \kms), which yields an average outflow rate $\dot{M}\simeq$25-30~\msyr. 





\section{Star formation properties of \tcd}
\label{secsfe}

\citet{Nesvadba10} found that powerful radio galaxies show an  SFR surface density lower by 10 to 50 times
than normal star-forming galaxies with the same molecular gas surface density. They calculated the SFR using the 7.7~$\mu$m PAH feature emission of their sample, except for \object{3C~326~N}, where they also used the 70~$\mu$m continuum emission as an SFR estimator  \citep{Ogle07}.  \citet{Nesvadba10} suggested that the low SFR in radio galaxies is due to a lower star-forming efficiency ($SFE$=$SFR$/\mhd) than in normal galaxies. Large-scale shocks in the ISM of radio galaxies may increase the turbulence of molecular gas \citep[see also][]{Nesvadba11}, which inhibits star formation to a large extent. 


In Sect. \ref{sfrestim}, we calculate the SFR of \tcd\ using all available estimators. Some SFR tracers are sensitive to AGN radiation and jet-induced shocks in the ISM, because these may increase the flux of ionized gas emission lines, vary the shape {of the continuum}, and destroy ISM molecules such as the PAH. Hence, we also discuss the possible caveats of the different SFR calibrations to find the most accurate value of the SFR for \tcd. In Sect. \ref{sec_sfe},  we study the SFE and the location of \tcd\ in the Kennicut-Schmidt (KS) diagram and discuss the differences with respect to the results obtained by \citet{Nesvadba10}. We complete the comparison sample with the young radio sources  \object{4C~12.50}, \object{4C~31.04} \citep{Willett10}, and the reactivated radio source \object{3C~236} \citep{Tremblay10, Labiano13}.


\subsection{Star formation rate}
\label{sfrestim}




\subsubsection{Near-ultraviolet photometry}
\label{photom}

\begin{figure*}
\centering
\includegraphics[width=1.8\columnwidth]{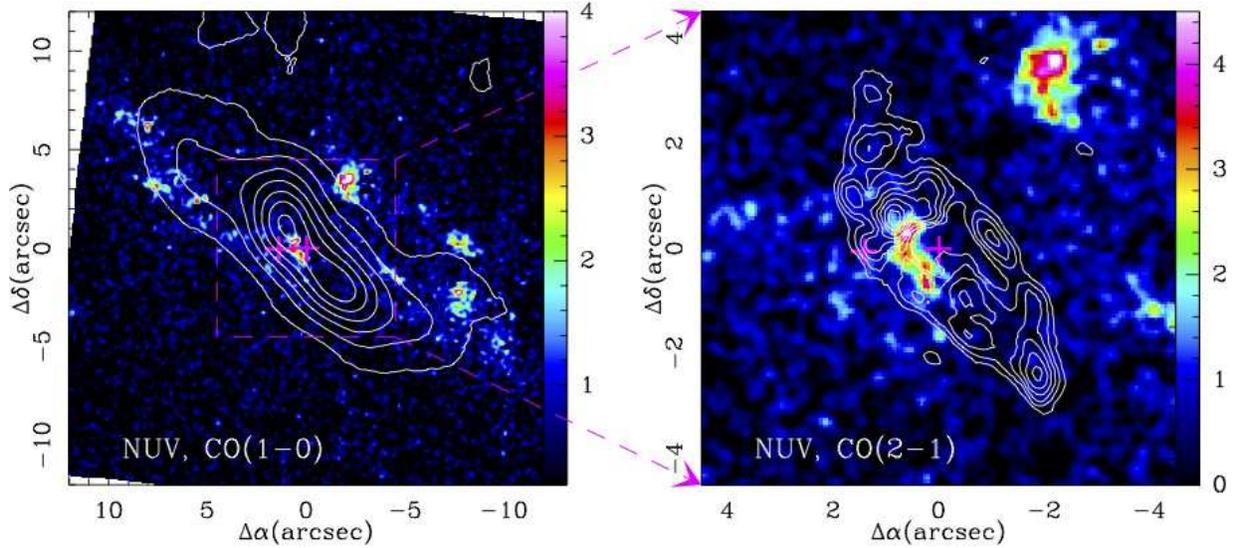} 
\caption{HST NUV image of \tcd, with the \couc\ (left) and \codu\ (right) emission overlaid. Contour levels as in Fig. \ref{cont_coemis}. Color version available in electronic format.  HST image in counts s\mone. \label{nuvoverlay}}
\end{figure*}

Figure \ref{nuvoverlay} shows an overlay of the HST-NUV image and the CO maps of \tcd. {  The morphology of the UV emission, which shows a highly clumpy distribution, indicates that most of the UV knots are associated with star formation \citep{Allen08, Baldi08}.  Inspection of Fig. \ref{nuvoverlay} suggests that the star-forming regions are  associated with the CO disk. The heavy obscuration seen in the $R-H$ map, and the patchy morphology of the NUV emission suggest that any SFR estimations based on the NUV image will be largely underestimated, however, and we therefore preferred { not to use it} to calculate the SFR.}

\begin{table}[t]
\caption{SFR estimations of \tcd.}
\label{tab_sfr}
\begin{minipage}{\columnwidth}
\centering
\resizebox{\textwidth}{!}{
\begin{tabular}{cccccccc} 
\hline
\hline
Region & Tracer &  SFR    & \mhd\  & Area  &  Refs. \\                 
       &        &   (\msun\ yr\mone) & (10$^9$ \msun)& (kpc$^2$) &\\
\hline
Unresolved & 24 $\mu$m   & 3.0 & 22 & 460 & 1,2\\
Nucleus & 24$\mu$m+H$\alpha$ & 3.2$\pm$0.1 & 3.3 & 5.4$^a$ & 1,2,3\\
Unresolved & 70 $\mu$m   & 5.9$\pm$0.1 & 22 & 460 & 1,4\\
Nucleus & 11.3 $\mu$m  & 2.5$\pm$0.2 & 8.2 & 28.3 & 5,6 \\
Nucleus & 6.2+11.3 $\mu$m  & 18$\pm$1 & 8.2 & 28.3 & 5,7,10\\
Nucleus & [\ion{Ne}{ii}] & 4.4$\pm$1 & 8.2 & 28.3 & 5,6\\
Nucleus & [\ion{Ne}{ii}]+[\ion{Ne}{iii}] & 11$\pm$1 & 8.2 & 28.3 & 5,7,10\\
Region 1 &  8\mum,dust & {  $<$}3.3 & 8.0 & 21.5 &  \ref{appregions} \\
Region 2 & 8\mum,dust   & 1.4 &1.9 & 19.7 &  \ref{appregions} \\
Region 3 &  8\mum,dust & 1.0 & 2.2 & 30.0 &  \ref{appregions} \\
Region 4 &  8\mum,dust  & 1.7 & 4.8 & 51.8 &  \ref{appregions} \\
Region 5 &  8\mum,dust & 0.5 & 2.5 & 13.5 &  \ref{appregions} \\
Region 6 & 8\mum,dust  & 0.5 & 1.1 & 9.3 &  \ref{appregions} \\
\hline
\end{tabular}
}
\end{minipage}
\\ \\
$^a$ The area of the 24 \mum\ measurement corresponds to the unresolved galaxy  (460 kpc$^2$).\\
Limits are 3$\sigma$. Uncertainties in \mhd\ are $\sim$$5\%$. SFR uncertainties listed when available.\\
References:
1-\citet{Dicken10}, 2-\citet{Calzetti07},3-\citet{Buttiglione09}, 4-\citet{Seymour11}, 5-\citet{Guillard12}, 6- \citet{Diamond12}, 7-\citet{Willett10},
8-\citet{Dicken11},  \ref{appregions}-Appendix \ref{appregions} of this work.
 \end{table}

\subsubsection{Optical spectroscopy}
\label{halfa}

The optical spectra available for \tcd\ are from the central 3$\arcsec$ of the source. In this region we can expect a large contribution from the AGN to the emission lines used to trace the SFR \citep[e.g., H$\alpha$ and \ion{O}{ii},][]{Kennicutt98}, which would cause an overestimated SFR value. 
On the other hand, the nuclear region of \tcd\  is crossed by large, thick, dust lanes, which absorb a large amount of optical emission-line flux. Thus, any SFR evaluation based on optical fluxes will be underestimated. The combination of both effects (overestimation due to the AGN emission, and underestimation due to dust obscuration) makes it impossible to calculate the contribution from star-forming regions to the optical emission line fluxes in the nucleus of \tcd. 




\subsubsection{Mid-infrared continuum}

\citet{Calzetti07}  showed that the 24 $\mu$m luminosity of a galaxy is indicative of its SFR \citep[see also][]{Rieke09, Kennicutt09, Calzetti10}. They presented two SFR calibrations: based on the  24 $\mu$m luminosity alone, and combined with the H$\alpha$ luminosity. If we apply these calibrations to the 24 $\mu$m  emission of \tcd\ \citep[$F_{24\mu\mathrm{m}}$=3.88$\times$$10^{-12}$ erg s\mone\ cm\mtwo,][]{Dicken10}, and the H$\alpha$ emission of \tcd\ \citep[$F_{\mathrm{H}\alpha}$=9.83$\times$10$^{-15}$erg s\mone\ cm\mtwo; SDSS,][]{Buttiglione10}, we obtain an $SFR$=3.0 \msun\ yr\mone\ and an $SFR$=3.2$\pm$0.1 \msun\ yr\mone for the 24$\mu$m and 24$\mu$m+H$\alpha,$ respectively. 
A possible caveat of the 24  \mum\ emission as SFR estimator is that this emission can be increased by the heating of dust by the AGN \citep{Tadhunter07}, yielding an overestimated SFR. However, \citet{Leipski09} found that the Spitzer IRS spectrum of \tcd\ is dominated by star formation, and the shape of the continuum is similar to the continuum of local star-forming {  galaxies \citep[e.g.,][]{Smith07}. Therefore the} SFR obtained from the 24  \mum\ emission is a good estimator of the global SFR of the obscured star-forming regions in \tcd.

The 70  \mum\ luminosity can be used to establish lower and upper limits on the SFR \citep{Seymour11,Symeonidis08, Kennicutt98b}. 
  The lower limit is obtained by subtracting the AGN contribution from the 70  \mum\ luminosity with the [\ion{O}{iii}] and 70 \mum\ correlation by \citet{Dicken10}. The upper limit is obtained assuming that all the 70  \mum\ emission is due to star formation.
Using the 70 $\mu$m flux of \tcd\ measured by  \citet{Dicken10} with {\it Spitzer} ($F_{70\mu\mathrm{m}}$=1.29$\times$$10^{-11}$ erg s\mone\ cm\mtwo), we obtained 5.8$<$SFR$<$6.0\msyr. The 70 $\mu$m emission can be increased by the heating of dust by the AGN \citep{Tadhunter07}, which would give an overestimated SFR. For \tcd, \citet{Dicken10} found that the 70 $\mu$m emission is $\sim$35 times brighter than expected for a nonstarburst galaxy with the same [\ion{O}{iii}] luminosity. The authors showed that this difference {arises because} star formation is the main contributor to the 70 $\mu$m flux. 
The SFR derived from 70 \mum\ for \tcd\ is consistent with the SFR derived from the 24 \mum\ emission, which supports the assumption that the contribution from the AGN to these fluxes is low, and the SFR estimations from the IR continuum are accurate.



\subsubsection{Mid-infrared spectroscopy}

Based on a sample of 25 Seyfert galaxies, \citet{Diamond10}  found that the [\ion{Ne}{ii}] emission line is a good tracer of the SFR in AGN. \citet{Willett10} used the [\ion{Ne}{iii}]$\lambda$15.6 $\mu$m plus the [\ion{Ne}{ii}]$\lambda$12.8 $\mu$m fluxes to measure the SFR in a sample of compact symmetric objects \citep[CSO, see also][]{Ho07}.  The neon emission lines of \tcd\ have fluxes $F_{[\ion{Ne}{ii}]}$ = (4.07$\pm$1.29)$\times$$10^{10}$ erg s\mone\ cm\mtwo\  and $F_{[\ion{Ne}{iii}]}$ = (1.27$\pm$0.09)$\times$$10^{10}$~erg~s\mone~cm\mtwo\ \citep[{\it Spitzer} spectrum,][]{Guillard12}, which yield  $SFR$ = 4.4$\pm$1~\msun~yr\mone\ and $SFR$ = 11$\pm$1~\msun~yr\mone for the [\ion{Ne}{ii}] and [\ion{Ne}{ii}]+[\ion{Ne}{iii}] relations, respectively.

The AGN contribution to the [\ion{Ne}{ii}] line luminosity can be estimated with the [\ion{O}{iv}]$\lambda$25.9 $\mu$m / [\ion{Ne}{ii}] $\lambda$12.8 $\mu$m or the [\ion{Ne}{v}]$\lambda$14.3 $\mu$m / [\ion{Ne}{ii}]$\lambda$12.8 $\mu$m ratios \citep{Diamond10, Sturm02, Genzel98}. For \tcd, the [\ion{O}{iv}]/[\ion{Ne}{ii}] and [\ion{Ne}{v}]/[\ion{Ne}{ii}] ratios are 0.07 and 0.04, respectively. Both values yield an AGN contribution below 10\% to the [\ion{Ne}{ii}] line (90\% starburst contribution), consistent with the results obtained by \citet{Leipski09} using the whole Spitzer IRS spectrum of \tcd. The higher SFR from the combined neon lines is probably due to contributions from the AGN to the [\ion{Ne}{iii}] line emission \citep{LaMassa12}. 
Another estimate of the AGN contribution is given by the ratio [\ion{Ne}{iii}]/[\ion{Ne}{ii}], which increases with the hardness of the ionizing environment. For \tcd, log [\ion{Ne}{iii}]/[\ion{Ne}{ii}]=--0.5, similar to the  mean value of the ratio in ULIRG and starburst galaxies (--0.35), and below the mean value for AGN (--0.07) and CSO \citep[--0.16,][]{Willett10}, suggesting a low contribution from the AGN in \tcd. \citet{Pereira10} studied the IR emission line ratios in a large sample of active and \ion{H}{ii} galaxies. Based on their results, the [\ion{O}{iv}]/[\ion{Ne}{ii}] and [\ion{Ne}{iii}]/[\ion{Ne}{ii}] ratios of \tcd\ are closer to those of LINER-like and HII galaxies than Seyferts or quasi-stellar objects. 
Therefore, the flux of the [\ion{Ne}{ii}] line of \tcd\ has negligible contributions from the AGN, and it is a reliable SFR estimator.

\subsubsection{PAH emission}
\label{sfrpah}

The mid-IR spectrum of \tcd\ shows clear emission lines from several PAH features: $F_{11\mu\mathrm{m}}$=(2.15$\pm$0.06)$\times$10$^{-13}$ erg s\mone\ cm\mtwo, $F_{7.7\mu\mathrm{m}}$=(5.57$\pm$0.25)$\times$10$^{-13}$ erg s\mone\ cm\mtwo\ and $F_{6.2\mu\mathrm{m}}$=(1.11$\pm$0.06)$\times$10$^{-13}$ erg s\mone\ cm\mtwo\ \citep[][]{Dicken11, Guillard12}. 

Based on the correlation found between the emission of the neon line and the 6.2 $\mu$m plus 11.3 $\mu$m PAH, \citet{Willett10} used the luminosities of these PAH to calculate the SFR in their sample \citep[see also][]{Farrah07}.  Applying this method to \tcd\ ($F_{6.3+11.3\mu\mathrm{m}}$=(3.26$\pm$0.12)$\times$$10^{-13}$ erg s\mone\ cm\mtwo), we obtained  SFR=18$\pm$1 \msyr. 

\begin{figure*}
\centering
\includegraphics[width=1.8\columnwidth]{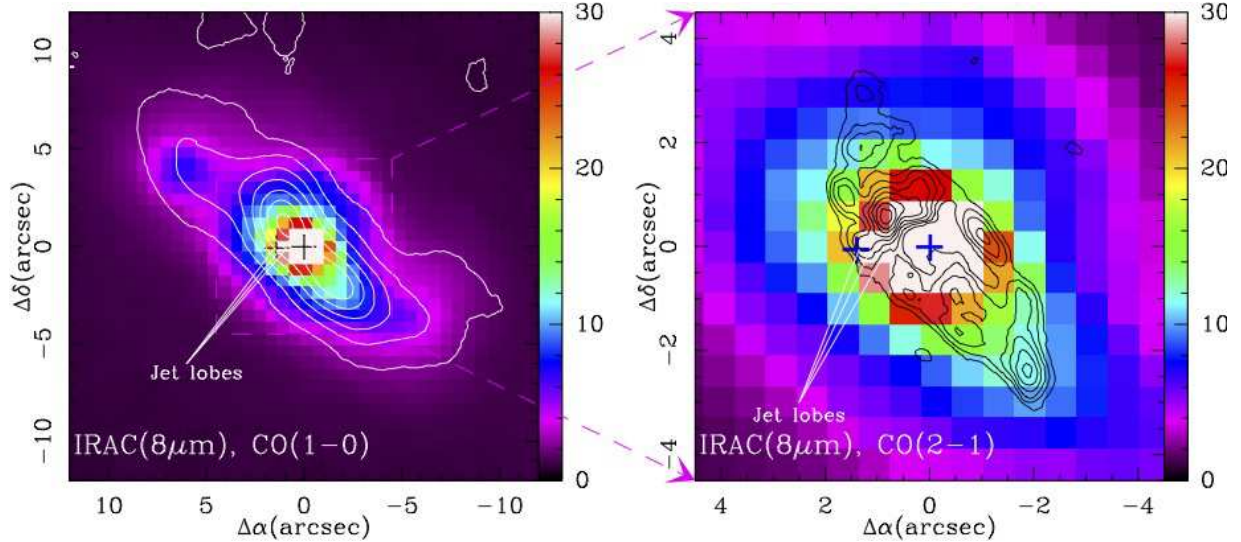}  
\caption{Spitzer/IRAC 8 \mum\ map of \tcd, with the \couc\ and \codu\ emission overlaid. Contour levels as in Fig. \ref{cont_coemis}. Color version available in electronic format. \label{iroverlay}}
\end{figure*}

By combining Eqs. 2, 3 and 8 of \citet{Calzetti07}, a relation between \so\ and SFR can be derived \citep[see e.g.,][]{Nesvadba10}; where \so\ is the luminosity surface density in the IRAC 8  \mum\ image with the stellar continuum removed using the IRAC 3.6 \mum\ image \citep{Helou04, Calzetti05}. Both the 8  \mum\ and 3.6  \mum\ IRAC images of \tcd\ are available in the Spitzer archive. We downloaded them and obtained an \so\ image\footnote{For our data, the variations in the f$_{3.6}$ factor of the correction \citep[0.22-0.29][]{Helou04} do not affect the results noticeably.}. 
{ To estimate the SFR, we divided \tcd\ into six different regions based on its NUV and CO morphology (the details can be found in Appendix \ref{appregions})  
and derived the SFR in each region. The results are summarized in Table \ref{tab_sfr}.} 
Our SFR estimates from \so\ are on average $\sim$ six times higher than the value obtained by \citet{Nesvadba10}. 
 {   \citet{Nesvadba10} calculated the SFR using the flux of the  7.7 \mum\ PAH line from the {\it Spitzer} nuclear spectrum of \tcd\ \citep{Ogle10}
 as \so\ in the \citet{Calzetti07} equations. 
The flux of the the  7.7 \mum\ PAH line was measured using PAHFIT \citep{Smith07}.  It should be noted that this flux estimate may be off by a factor $\lesssim$2 due to the extended PAH emission of \tcd\ \citep{Ogle10}, while the IRAC image includes the whole extension of the PAH emission.
 In our analysis, we used the complete bandpass of the 8 \mum\ IRAC image (6.4 to 9.3  \mum, Fig. \ref{iroverlay}), corrected for the stellar continuum, as done by \citet{Calzetti07}. } 
  Inspection of the MIR spectrum of \tcd\ shows that the 7.7 \mum\ PAH line emission represents only $\lesssim$15\% of the nuclear flux in the 8 \mum\ IRAC band \citep[e.g.,][]{Ogle10, Guillard12}.  Therefore, the origin the discrepancy with the \citet{Nesvadba10} results seems to be the different fluxes considered for \so. Using the 7.7 \mum\ PAH flux as \so\ underestimates the SFR of \tcd.

\citet{Diamond10} found that the 11.3  \mum\ PAH emission, by itself, is a reliable estimator of the SFR in Seyfert galaxies, unlike the 6.3 and 7.7 \mum\ PAH \citep[see also][]{Smith07, Diamond12, Esquej14}. They found a strong correlation between the emission of the  [\ion{Ne}{ii}] and 11.3  \mum\ lines in their sample due to the star formation of the host galaxies, with no contributions from the AGN. By comparing the [\ion{Ne}{ii}] and 11.3  \mum\  PAH emission of \tcd\ with their sample (Fig. 5 of their paper), it is clear that the emission of \tcd\ is consistent with that correlation\footnote{\citet{LaMassa12} found that the 11.3 \mum\ PAH is significantly suppressed in AGN-dominated systems. However, this may not apply to \tcd, because it is a starburst-dominated system \citep{Dicken10, Leipski09}.}. Using the 11.3 \mum\ emission of \tcd, we obtain a reliable estimation for the SFR of \tcd: SFR=2.5$\pm$0.2\msyr. 

\subsection{Star formation efficiency}
\label{sec_sfe}

\begin{figure}
\centering
\includegraphics[width=\columnwidth]{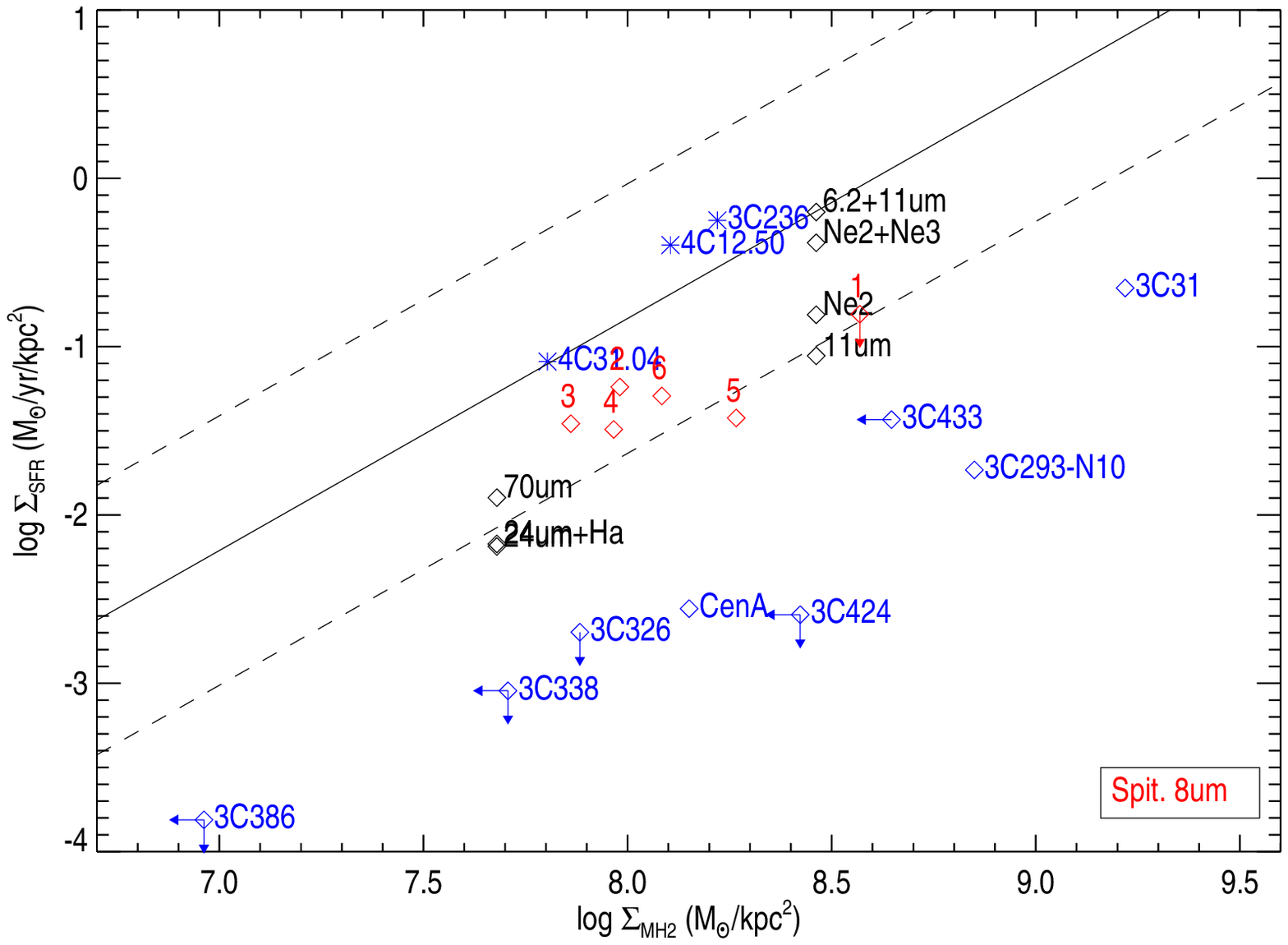} 
\caption{{  \ssfr\ and \shd\ of \tcd\ (black and red), the sample
of \citet{Nesvadba10}  (blue diamonds), young sources  \object{4C~12.50}, \object{4C~31.04}, and \object{3C~236} (blue stars). Red diamonds show the Spitzer \so\ data for \tcd.} Black diamonds show the SFR estimates for \tcd\ using the rest of tracers discussed in the text. DRNe2 and DR11 show the SFR estimates using the recipes
of \citet{Diamond12}, DS70 is the SFR estimate based on the method
of \citet{Seymour11}, \citet{Dicken10}.
  Solid line: best-fit of the KS-law from \citet{Kennicutt98}. Dashed lines: dispersion around the KS-law best fit for normal star-forming galaxies \citep{Roussel07, Kennicutt98}. \label{kslaw}}
\end{figure}

As discussed above, the most accurate SFR tracers for \tcd\ are the 11.3 \mum\ PAH, and [\ion{Ne}{ii}] emission lines and the 24 \mum, 70 \mum\ continuum emission, which yield an average SFR of  4.0$\pm$1.5 \msyr.  Therefore, $SFE_{\rm{3C293}}$=0.18~Gyr\mone, and $t_{\rm{deplete}}$$\sim$5.6 Gyr.
The average mass rate of the \hi\ outflow is $\dot{M}\simeq$25-30~\msyr\ (Sect \ref{coldmg}), suggesting that the outflow might deplete the gas reservoir faster than the star formation, lowering the SFE of \tcd.
Figure \ref{kslaw} compares the canonical KS-law fitted for normal star-forming galaxies \citep{Roussel07, Kennicutt98} with the SFR surface density (\ssfr=$SFR$/$area$) and cold-\hd\ mass  surface density (\shd=$M(H_2)$/$area$) for \tcd,  the young sources  \object{4C~12.50}, \object{4C~31.04} \citep{Willett10}, the re-started source \object{3C~236} \citep{Labiano13}, and the
sample of \citet{Nesvadba10}\footnote{The data from \citet{Nesvadba10} did not include warm-\hd. Therefore, we considered only the cold-\hd\ mass for the comparison of star-formation laws, which yields log \shd$\simeq$7.7 \msun~kpc\mtwo. Adding the cold and warm-\hd\ masses of \tcd\ yields log \shd$\simeq$7.9~\msun~kpc\mtwo.}. 
Inspection of Fig. \ref{kslaw} shows that \tcd\ is a normal-efficiency (consistent with the KS-law) star-forming radio galaxy and is 10--50 times more efficient than the sample of \citet{Nesvadba10}.  This means that the outflow seen in \hi\ and ionized gas does not seem to affect the molecular-gas-forming stars. 

The \ssfr\ given by \citet{Nesvadba10} for \tcd\ (labeled \tcd-N10 in Fig. \ref{kslaw}) is lower by a factor 2.5 than our \ssfr\ estimates using \so, the 11 \mum\ PAH and the [\ion{Ne}{ii}] emission line. 
Inspection of Fig. \ref{kslaw} shows that increasing the \ssfr\ of \tcd-N10 by a 2.5 factor does not make it consistent with the KS-law, however. The \shd\ would be too large compared with our results. \citet{Nesvadba10} used the \mhd\ data published by \citet{Evans99b} to calculate the column density of \hd\ in \tcd. Combining their single-dish and 3.5\arcsec-resolution map of the \couc\ emission, \citet{Evans99b} argued that the 10$^{10}$ \msun\ of \hd\ are distributed along a $\sim$7\arcsec\ (6 kpc) diameter disk, yielding a surface density \shd=7.1$\times$10$^8$ \msyr/kpc\mtwo. Our higher resolution and sensitivity \couc\ map shows that the molecular gas is distributed along a $\sim$24$\arcsec$ (21~kpc-) diameter disk, with an average density of \shd=6.4$\times$10$^8$ \msyr\ kpc\mtwo. Applying the latter density to \tcd-N10 gives an SFE consistent with the KS-law.

Our previous study of feedback in radio galaxies \citep{Labiano13} showed that the young ($\lesssim$$10^5$ yr) radio sources \object{3C~236},  \object{4C~12.50}, and \object{4C~31.04} had an unexpectedly high SFE compared with the evolved ($\gtrsim$$10^{7}$ yr) radio galaxies in the sample of  \citet{Nesvadba10}. Based on the age of these sources, we proposed an evolutionary scenario where the impact of AGN feedback on the ISM of the host galaxy would increase with the age of the radio source. Hence, older radio sources would show lower SFE than young sources. 
The large radio jets of \tcd\ have been active for the past $\sim$2$\times$10$^7$ yr, which gives enough time for the feedback effects (such as an outflow) to spread across the ISM and suppress the star formation in the host. The SFE of \tcd\ is consistent with that of normal star-forming galaxies and young radio sources, with no evidence of quenched star formation, however. Therefore, the evolutionary scenario proposed in \citet{Labiano13} does not seem to apply for \tcd.
Moreover, a recent study of the SED (from UV to radio wavelengths) of radio-loud AGN found that the SFR in these systems is mildly diminished by the radio jets, but similar (or higher) to the SFR of inactive galaxies \citep{Karouzos13}. 
 Our results suggest that the apparently low SFE found in evolved radio galaxies is probably caused by an underestimation of the SFR and/or an overestimation of the \hd\ density and is not a property of these sources. 

\section{Summary and conclusions}


We carried out 1~mm, 3~mm continuum and \couc, \codu\ line PdBI observations of the powerful radio galaxy \tcd. The host of \tcd\ was known to harbor active star-forming regions and outflows on \hi\ and ionized gas. The  high sensitivity and resolution of our data in combination with HST and Spitzer images and \hi\ spectra allowed us to study the impact of AGN feedback on the host of this evolved radio galaxy in detail. We investigated the kinematics and distribution of the molecular gas to search for evidence of outflow motions. We also accurately determined the SFE of the host galaxy and compared it with the efficiencies of normal star-forming galaxies and young and old radio sources. The main results of our work are summarized below:

\begin{itemize}

\item 

{ 
The 1~mm and 3~mm maps separate the continuum emission from the AGN and the eastern jet of \tcd. The emission of the AGN and jet at 1~mm and 3~mm are consistent with synchrotron radiation.
}

\item

{  The cold molecular gas of \tcd, traced by the \couc\ emission, forms a disk with a diameter of 21~kpc, with a mass \mhd=2.2$\times$10$^{10}$ \msun. The dust mapped by the HST $R$--$H$ image and the star formation seen in the HST-NUV and Spitzer-8 \mum\ images are clearly associated with the molecular gas disk. 
The morphology of the CO emission is consistent with a warped, corrugated disk, with the outer edges ($R\gtrsim8$ kpc) seen
almost edge-on, and the central ($R\lesssim4$ kpc) region tilted toward the plane of the sky, with the southern side facing us.
}

\item

{  The kinematics of the CO disk is consistent with rotation around the core, with no evidence of fast ($\gtrsim$500 \kms) outflowing molecular gas. 
 The detailed analysis of the CO kinematics showed that the inclination and the position angle of the disk are consistent with a corrugated, warped morphology. Furthermore, the Fourier decomposition of the velocity field {is consistent with} the presence of a warp in the CO disk of \tcd, as suggested by the CO maps.
Using the kinematics of the \couc\ emission line, we determined the systemic velocity of \tcd\ to be \vsco=13\,434$\pm$8~\kms.  
}

\item

The CO spectra of \tcd\ reveal several CO absorption features toward the core and jet coordinates. These absorptions are consistent with those observed in the \hi\ spectra, which originate in the disk. 
 The comparison of the CO and \hi\ spectra also showed that the velocities of the \hi\ outflow that cannot be explained by rotation of the disk range from $\sim$300 \kms\ to $\sim$900 \kms\ with respect to the systemic velocity.

\item

We studied the different SFR tracers available for \tcd, from NUV to IR, to obtain an accurate estimate of its SFR. The most reliable tracers are the [\ion{Ne}{ii}] and 11.3 \mum\ PAH emission, as well as the 24 and 70 \mum\ continuum, which, for \tcd, have negligible AGN contributions. The average SFR value for \tcd\ is 4.0$\pm$1.5 \msyr.

\item
The SFE of \tcd\ ($SFE$=0.18~Gyr\mone) is consistent with the KS-law of normal galaxies and young radio galaxies. It is 10-50 times higher than the efficiencies measured in other evolved radio galaxies. Therefore, the evolutionary scenario of AGN feedback proposed in our previous work \citep{Labiano13}  does not apply to the evolved radio galaxy \tcd. The higher SFE of \tcd\ is a global property of the molecular gas disk.


\item 
The origin of the discrepancy between the SFE of powerful radio galaxies and normal, KS-law, star-forming galaxies, might be due to the use of the 7.7\mum\ PAH emission, which probably underestimates the SFR in powerful AGN environments, and/or an underestimation of the \hd\ surface densities in radio galaxies. 

\end{itemize}

\begin{acknowledgements}

We are grateful to A. Alonso-Herrero, P. Esquej, N. Nesvadba, E. Bellocchi and K. Dasyra for fruitful scientific discussions. We also thank the anonymous referee for very useful comments and suggestions. AL acknowledges support by the Spanish MICINN within the program CONSOLIDER INGENIO 2010, under grant ASTROMOL (CSD2009-00038), Springer and EAS. This research has made use of NASA's Astrophysics Data System Bibliographic Services and of the NASA/IPAC Extragalactic Database (NED) which is operated by the Jet Propulsion Laboratory, California Institute of Technology, under contract with the National Aeronautics and Space Administration. Based on observations made with the NASA/ESA Hubble Space Telescope, and obtained from the Hubble Legacy Archive, which is a collaboration between the Space Telescope Science Institute (STScI/NASA), the Space Telescope European Coordinating Facility (ST-ECF/ESA) and the Canadian Astronomy Data Centre (CADC/NRC/CSA). Funding for the SDSS and SDSS-II has been provided by the Alfred P. Sloan Foundation, the Participating Institutions, the National Science Foundation, the United States Department of Energy, NASA, the Japanese Monbukagakusho, the Max-Planck Society, and the Higher Education Funding Council for England. The SDSS website is http://www.sdss.org/. The SDSS is managed by the Astrophysical Research Consortium for the Participating Institutions. The Participating Institutions are the American Museum of Natural History, Astrophysical Institute Potsdam, University of Basel, University of Cambridge, Case Western Reserve University, University of Chicago, Drexel University, Fermilab, the Institute for Advanced Study, the Japan Participation Group, Johns Hopkins University, the Joint Institute for Nuclear Astrophysics, the Kavli Institute for Particle Astrophysics and Cosmology, the Korean Scientist Group, the Chinese Academy of Sciences (LAMOST), Los Alamos National Laboratory, the Max-Planck Institute for Astronomy (MPIA), the Max-Planck Institute for Astrophysics (MPA), New Mexico State University, Ohio State University, University of Pittsburgh, University of Portsmouth, Princeton University, the United States Naval Observatory and the University of Washington. This work makes use of euro-vo software, tools or services and TOPCAT \citep{Taylor05}. Euro-vo has been funded by the European Commission through contract numbers RI031675 (DCA) and 011892 (VO-TECH) under the Sixth Framework Programme and contract number 212104 (AIDA) under the Seventh Framework Programme.

\end{acknowledgements}

\bibliographystyle{aa}

\appendix
\section{Spatially resolved SFE of \tcd}
\label{appregions}

\begin{figure}[b]
\centering
\includegraphics[width=\columnwidth]{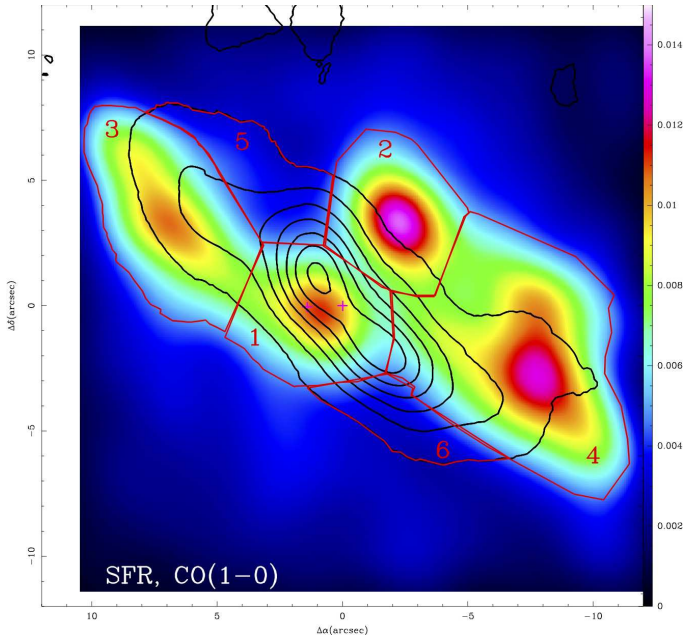}
\caption{Overlay of the NUV image and \couc\ map of \tcd, showing the different regions for the analysis of the SFR and \hd. The NUV image has been smoothed with the beam of CO, and the NUV flux has been transformed into SFR using Eq. 1 of \citet{Kennicutt98b}. Contour levels as in Fig. \ref{rhoverlay}. Color version available in electronic format.  \label{regions}}
\end{figure}

The NUV and 8  \mum\ emission of \tcd\ are almost co-spatial to the \codu\ disk, suggesting that \tcd\ is using the cold molecular gas traced by CO to form stars. The spatial resolution of the HST image and the CO map are sufficient to study the SFE along the molecular gas disk of \tcd. In this appendix, we describe how we measured the SFR and SFE in different sections of the molecular gas disk.

{  Owing to its higher resolution, we used the NUV image to define six `star-forming' regions in \tcd\ (although it was not used to map the resolved SFR because of {its sensitivity to obscuration, Sect. \ref{photom}).}  
Regions 1 to 4 were defined to include all NUV emission over three times{  the signal-to-noise threshold.}} 
 Regions 5 and 6 are the areas of the \couc\ disk where the NUV emission is below that level.
 Region 1 contains the NUV emission from the core. Region 2 contains the NW emission, which originates at the bottom of the large NW radio jet. Regions 3 and 4 contain the NUV emission NE and SW of the AGN. Regions 5 and 6 are defined by the remaining areas of the molecular gas disk that are not included in regions 1 through 4. { Figure \ref{regions} shows the location of the different regions on the CO disk on top of the NUV image, smoothed to match the resolution of the \couc\ data} .


{  
We corrected the 8 $\mu$m emission from the stellar continuum using the 3 $\mu$m map \citep{Helou04, Pahre04, Calzetti05} to obtain \so. 
We then integrated \so\ in each region and applied the equations in \citet{Calzetti07} to estimate their SFR. These estimates and the \hd\ contents for each region are listed in Table \ref{tab_sfr}. We measured 
 $SFE_{\rm{R1}}$$\sim$0.4 Gyr\mone,  $SFE_{\rm{R2}}$$\sim$0.6 Gyr\mone,  $SFE_{\rm{R3}}$=0.5 Gyr\mone,  $SFE_{\rm{R4}}$=0.3 Gyr\mone, $SFE_{\rm{R5}}$=0.2 Gyr\mone,  and $SFE_{\rm{R6}}${ =}0.4 Gyr\mone. 
 The emission of regions 1 and 2 probably has contributions from the AGN and jet. The CO emission in region 1 is affected by absorption and the \mhd\ is a lower limit. Therefore the SFR value of region 1 is merely indicative. 
 }
 
  The SFE estimates for each section of the disk are consistent with the SFE of galaxies following the KS law \citep[e.g., SINGS sample,][]{Kennicutt03, Roussel07}. Therefore the higher SFE of \tcd, compared with the radio galaxies in \citet{Nesvadba10}, cannot be attributed to an individual region; \tcd\ shows an SFE consistent with the KS-law all along its molecular gas disk.

\end{document}